\documentclass[preprint,review,12pt]{elsarticle}

\usepackage{amssymb}
\usepackage{siunitx}
\usepackage{xcolor}
\usepackage{amsmath}
\usepackage{quiver}
\usepackage{booktabs}
\usepackage{framed}
\usepackage{multicol}
\usepackage{longtable}
\usepackage[stdsubgroups,nocfg]{nomencl}
\usepackage{setspace}
\usepackage{hyphenat}
\usepackage{import}
\usepackage{float}
\usepackage{dblfloatfix}
\usepackage{gensymb}
\usepackage{hyperref}

\renewcommand\nomgroup[1]{%
	\item[\bfseries
	\ifstrequal{#1}{A}{}{%
	\ifstrequal{#1}{G}{Greek Letters}{%
	\ifstrequal{#1}{S}{Subscripts}{%
	\ifstrequal{#1}{Z}{Acronym}{%
	}}}}%
]}

\newcommand\numberthis{\addtocounter{equation}{1}\tag{\theequation}}

\makenomenclature

\usepackage{lineno}

\usepackage{tikz}
\usepackage{tkz-euclide}

\usepackage[acronym]{glossaries}

\newcommand{\nw}{{n_{\mathrm{w}}}}
\newcommand{\ns}{{n_{\mathrm{s}}}}

\newcommand{\f}{\mathrm{f}}
\newcommand{\tu}{\mathrm{p}}
\newcommand{\s}{\mathrm{s}}

\newcommand{\w}{\mathrm{w}}
\newcommand{\hx}{\mathrm{hx}}
\newcommand{\h}{\mathrm{h}}
\newcommand{\n}{\mathrm{n}}
\newcommand{\nc}{\mathrm{nc}}
\newcommand{\co}{\mathrm{c}}
\newcommand{\is}{\mathrm{is}}
\newcommand{\p}{\mathrm{prec}}
\newcommand{\fus}{\mathrm{fus}}
\newcommand{\ice}{\mathrm{ice}}
\newcommand{\hhx}{\mathrm{hhx}}
\newcommand{\supply}{\mathrm{sup}}
\newcommand{\ret}{\mathrm{ret}}
\newcommand{\bp}{\mathrm{bp}}
\newcommand{\bo}{\mathrm{b}}
\newcommand{\dr}{\mathrm{d}}
\newcommand{\Sim}{\mathrm{sim}}

\newcommand{\Rey}{\mathrm{Re}}
\newcommand{\dn}{\mathrm{d}}
\newcommand{\dt}{\mathrm{d}t}
\newcommand{\m}{\mathrm{m}}
\newcommand{\coTWO}{$\mathrm{CO}_{2}\mathrm{eq}$}
\newcommand{\adj}{\mathrm{a}}
\newcommand{\out}{\mathrm{o}}

\newacronym{5GDHC}{5GDHC}{fifth generation district heating and cooling}
\newacronym{sas}{SAs}{solar absorbers}
\newacronym{hps}{HPs}{heat pumps}
\newacronym{hp}{HP}{heat pump}
\newacronym{dhw}{DHW}{domestic hot water}
\newacronym{ltes}{LTES}{latent thermal energy storage}
\newacronym{sh}{SH}{space heating}
\newacronym{ode}{ODE}{ordinary differential equation}
\newacronym{odes}{ODEs}{ordinary differential equations}
\newacronym{acm}{ACM}{adsorption cooling machine}
\newacronym{ac}{AC}{adsorption cooling}
\newacronym{fc}{FC}{free cooling}
\newacronym{hts}{HTS}{high temperature storage}
\newacronym{try}{TRY}{test reference year}
\newacronym{eu}{EU}{European Union}
\newacronym{cvrmse}{CVRMSE}{coefficient of the variation of the root mean square error}
\newacronym{nmbe}{NMBE}{normalized mean bias error}
\newacronym{DIN}{DIN}{German Institute for Standardisation (de: Deutsches Institut für Normung)}
\newacronym{FMI}{FMI}{Functional Mock-up Interface}
\newacronym{ASHRAE}{ASHRAE}{American Society of Heating, Refrigerating and Air-Conditioning Engineers}

\journal{Applied Energy}

\begin{document}

\begin{frontmatter}

	\title{A detailed simulation model for \acrlong{5GDHC} networks with seasonal latent storage evaluated on field data}

	\author[first,fifth]{Manuel Kollmar\corref{cor1}}
	\cortext[cor1]{Corresponding author.}
	\ead{manuel.kollmar@h-ka.de}

	\author[second]{Adrian B{\"u}rger}
	\ead{adrian.buerger@pathtozero.de}

	\author[second]{Markus Bohlayer}
	\ead{markus.bohlayer@pathtozero.de}
	
	\author[third]{Angelika Altmann-Dieses}
	\ead{a.altmann-dieses@hs-mannheim.de}

	\author[first,fourth]{Marco Braun}
	\ead{marco.braun@h-ka.de}
	
	\author[fifth,sixth]{Moritz Diehl}
	\ead{moritz.diehl@imtek.uni-freiburg.de}

	\address[first]{Institute of Refrigeration, Air-Conditioning, and Environmental Engineering (IKKU), Karlsruhe University of Applied Sciences, Moltkestraße 30, 76133 Karlsruhe, Germany}
	\address[second]{Path to Zero GmbH, Haid-und-Neu-Straße 7, 76131 Karlsruhe, Germany}
	\address[third]{Technische Hochschule Mannheim, Paul-Wittsack-Straße 10, 68163 Mannheim, Germany}
	\address[fourth]{Faculty of Management Science and Engineering, Karlsruhe University of Applied Sciences, Moltkestraße 30, 76133 Karlsruhe, Germany}
	\address[fifth]{Systems Control and Optimization Laboratory, Department of Microsystems Engineering (IMTEK), University of Freiburg, Georges-Koehler-Allee 102, 79110 Freiburg im Breisgau, Germany}
	\address[sixth]{Department of Mathematics, University of Freiburg, Ernst-Zemelo-Stra{\ss}e 1, 79104 Freiburg im Breisgau, Germany}

	\begin{abstract}

		\Acrfull{5GDHC} networks accelerate the use of renewable energies in the heating sector and enable flexible, efficient and future-proof heating and cooling supply via a single network.  
		Due to their low temperature level and high integration of renewables, \acrshort{5GDHC} systems pose new challenges for the modeling of these networks in order to simulate and test operational strategies. 
		A particular feature is the use of uninsulated pipes, which allow energy exchange with the surrounding ground. 
		Accurate modeling of this interaction is essential for reliable simulation and optimization. 
		This paper presents a thermo-physical model of the pipe connections, the surrounding soil, a latent heat storage in the form of an ice storage as a seasonal heat storage and the house transfer stations. 
		The model is derived from mass and energy balances leading to \acrfull{odes}. 
		Validation is performed using field data from the \acrshort{5GDHC} network in Gutach-Bleibach, Germany, which supplies heating and cooling to 30 modern buildings. 
		With an average model deviation of \SI{4.5}{\percent} in the \acrfull{nmbe} and \SI{15.9}{\percent} in the \acrfull{cvrmse}, the model’s accuracy is validated against the available temperature measurements. 
		The realistic representation of the thermal-hydraulic interactions between soil and pipes, as well as the heat flow within the network, confirms the accuracy of the model and its applicability for the simulation of \acrshort{5GDHC} systems.
		The model is made openly accessible under an open-source license.		
	\end{abstract}

	\begin{keyword}

		\acrshort{5GDHC} network \sep Energy system model  \sep Seasonal energy storage \sep Heating network \sep Renewable energy

	\end{keyword}

\end{frontmatter}

\section{Introduction}

Decarbonizing the energy use of buildings is one of the biggest challenges of the ongoing energy transition.
Buildings account for \SI{30}{\percent} of the global energy consumption and \SI{26}{\percent} of CO\textsubscript{2} emissions, making them a key leverage factor \cite{IEA2023}.
In the \acrfull{eu}, about \SI{40}{\percent} of the final energy is used in buildings, accounting for \SI{36}{\percent} of CO\textsubscript{2} emissions  \cite{Economidou2020}.
Due to the high share of energy demand and the continued large dependence on fossil energy sources, ambitious targets have been set in recent years.
Germany, for example, set its goal to a reduction of carbon emissions of $\SI{65}{\percent}$ (in comparison to 1990) until 2030 \cite{BT2021}, which is ten percentage points above the target set by the \acrshort{eu} as part of the 'Fit-for-55' program \cite{EU2021}.
Based on the original version of the law and the level of emissions in the building sector in 2024, emissions would have to be reduced by 38 million tons of \coTWO\ to 67 million tons of \coTWO/a over the next six years \cite{BT2021,Agora2025}, while the Umweltbundesamt predicts a reduction gap of 110 million tonnes of \coTWO\ in the building sector up to 2030 \cite{UBA2025}.
The \acrshort{eu} Commission recently presented a recommendation for a $\SI{90}{\percent}$ reduction in greenhouse gases from 1990 levels by 2040 to pave the way for climate neutrality by 2050 \cite{EU2024}.
European emissions trading is also part of the 'Fit-for-55' program.
From 2027, buildings will be included in the second phase, which means that certificates will have to be bought for emitting CO\textsubscript{2} emissions, for example from burning fossil fuels to heat buildings.
The gradual reduction in certificates will likely result in higher CO\textsubscript{2} prices, making fossil fuels less economically viable.

\Acrfull{5GDHC} networks are one promising technology to help achieve these targets.
In contrast to previous generations of heating networks, \acrshort{5GDHC} networks operate with supply temperatures close to the ambient temperature in the range of $ \SIrange[range-phrase=-, range-units=single]{0}{30}{\degreeCelsius} $.
Decentralized \acrfull{hps} on the consumer side are used to raise the temperature of the heat transfer medium to the required level, given the low network temperature.
Due to the relatively low temperature level, \acrshort{5GDHC} networks usually offer the possibility of additional cooling, allowing heating and cooling to be provided via the same network.
In recent years, the number of \acrshort{5GDHC} networks built and planned has significantly increased.
Since 2011, at least 52 networks have been built or are planned to go into operation by 2026 in Germany \cite{Wirtz2022}.
Due to the relatively new concept and the much smaller number of such networks compared to conventional heating networks, there are few to no models of the typical components of a \acrshort{5GDHC} network that can be used to simulate or optimize such networks.
Existing models from previous heating network generations can only be used to a limited extent, as the fifth generation has different requirements for the models in certain cases, such as the detailed consideration of the ground when using uninsulated pipes.
Further, the networks often significantly differ in their design and operation strategies.
While the majority of the \acrshort{5GDHC} networks in Germany use geothermal energy as the main heat source \cite{Wirtz2022}, they can differ, among other things, in the installed energy systems for generation and consumption, the use of insulated or uninsulated pipes, a wide range of number of actors within the network, the network length and the network temperature itself.
Depending on the design of the grid, additional (seasonal) storage systems are used. 
For example, geothermal probes are often operated at seasonal temperatures or the grids are supplemented with sensible or latent heat storage systems for seasonal energy storage.

The network temperature is also crucial for the overall efficiency of the network, due to the prevalent use of \acrshort{hps} on the consumer side. 
With the common use of uninsulated pipes in this temperature range, the system temperatures also depend on the thermal interaction between the pipe system and the surrounding ground. 
These interactions and their influence on the optimal design and optimization of such systems require accurate models to capture these effects for use in planning and sizing of \acrshort{5GDHC} networks and in testing operational strategies.
\newline

\begin{supertabular}[t!]{p{0.125\textwidth} p{0.125\textwidth} p{0.75\textwidth}}
	Nomenclature & & \\
	\toprule
	Symbol & Unit & Description \\
	\midrule
	$ \alpha $ & $ \SI{}{\frac{\watt}{\meter\squared\kelvin}} $ & heat transfer coefficient \\
	$ A $ & $ \SI{}{\meter \squared} $ & area \\
	$ \beta $ & - & angle \\
	$ c $ & $ \SI{}{\frac{\joule}{\kilogram\kelvin}} $ & specific heat capacity \\
	$ C $ & $ \SI{}{\frac{\joule}{\meter\cubed\kelvin}} $ & volumetric heat capacity \\
	$ d $ & $ \SI{}{\meter} $ & distance \\
	$ \delta $ & $ \SI{}{\meter} $ & thickness \\
	$ \eta $ & - & efficiency \\
	$ f $ & - & Darcy friction factor \\
	$ \Delta h $ & $ \SI{}{\frac{\joule}{\kilogram\kelvin}} $ & specific enthalpy (normalized) \\
	$ l $ & $ \SI{}{\meter} $ & length \\
	$ k $ & $ - $ & correction factor \\
	$ \lambda $ & $ \SI{}{\frac{\watt}{\meter\kelvin}} $ & thermal conductivity \\
	$ m $ & $ \SI{}{\kilogram} $ & mass \\
	$ \dot{m} $ & $ \SI{}{\frac{\kilogram}{\second}} $ & mass flow \\
	$ \mu $ & $ \SI{}{\pascal\second}{} $ & dynamic viscosity \\
	$ n $ & - & number \\
	$ P $ & $ \SI{}{\watt} $ & power \\
	$ \Delta p $ & $ \SI{}{\pascal} $ & pressure drop \\
	$ \phi $ & - & fraction \\
	$ Q $ & $ \SI{}{\joule} $ & heat \\
	$ \dot{Q} $ & $ \SI{}{\watt} $ & heat flow \\
	$ r $ & $ \SI{}{\meter} $ & radius \\
	$ \Rey $ & - & Reynolds number \\
	$ \rho $ & $ \SI{}{\frac{\kilogram}{\meter\cubed}} $ & density \\
	$ s $ & $ \SI{}{\meter} $ & arc length \\
	$ t $ & $ \SI{}{\second} $ & time \\
	$ T $ & $ \SI{}{\degreeCelsius} $ & temperature \\
	$ \Delta T $ & $ \SI{}{\kelvin} $ & temperature difference \\
	$ U $ &  $ \SI{}{\frac{\watt}{\meter\squared\kelvin}} $  & thermal transmittance \\
	$ V $ & $ \SI{}{\meter\cubed} $ & volume \\
	$ w $ & - & water share \\
	$ y $ & - & position \\
	$ z $ & $ \SI{}{\meter} $ & height \\
	\bottomrule
	Subscript & & Description \\
	\midrule
	$ \adj $ & & adjacent \\
	$ \bo $ & & boundary \\
	$ \bp $ & & bypass \\
	$ \co $ & & concrete \\
	$ \mathrm{ch} $ & & chord \\			
	$ \dr $ & & dry \\
	$ \mathrm{el} $ & & electric \\
	$ \f $ & & fluid \\
	$ \fus $ & & fusion \\
	$ \hx $ & & heat exchanger \\
	$ \hhx $ & & household heat exchanger \\
	$ \h $ & & hollow \\
	$ \mathrm{ice} $ & & ice \\
	$ \is $ & & ice storage \\
	$ \m $ & & measured \\
	$ \mathrm{max} $ & & maximum \\
	$ \mathrm{min} $ & & minimum \\
	$ \n $ & & network \\
	$ \nc $ & & natural convection \\
	$ \out $ & & outer \\
	$ \tu $ & & pipe \\
	$ \p $ && precipitation \\
	$ \s $ & & soil \\
	$ \Sim $ & & simulated \\
	$ \supply $ & & supply \\
	$ \ret $ & & return \\
	$ \w $ & & water \\
	\bottomrule
	Acronym & & Description \\
	\midrule
	\acrshort{5GDHC} & & \acrlong{5GDHC} \\
	\acrshort{ASHRAE} & & \acrlong{ASHRAE} \\
	\acrshort{cvrmse} & & \acrlong{cvrmse} \\
	\acrshort{DIN} & & \acrlong{DIN} \\
	\acrshort{FMI} & & \acrlong{FMI} \\
	\acrshort{eu} & & \acrlong{eu} \\
	\acrshort{hps} & & \acrlong{hps} \\
	\acrshort{ltes} & & \acrlong{ltes} \\
	\acrshort{nmbe} & & \acrlong{nmbe} \\
	\acrshort{odes} & & \acrlong{odes} \\
	\acrshort{sas} & & \acrlong{sas} \\
	\bottomrule
\end{supertabular}

\subsection{Relevant literature}
\acrshort{5GDHC} networks have different modeling requirements compared to previous generations of heating networks. 
The integration of renewable energy makes technologies such as fossil burners and fossil fuels in general obsolete. 
In addition, the heating and cooling supply in the respective networks was designed for the highest and lowest temperature levels, while the latest generation generally requires additional heat producers at the building level.

Especially in the case of uninsulated pipes with low thermal resistance, the heat exchange between the underground pipe system and the ground must be taken into account \cite{Del_Hoyo_Arce2015}.
A thermal resistance model is proposed by \citet{Del_Hoyo_Arce2015} to integrate the transient heat gains or losses in the underground pipe system.
\citet{Hirsch2022} propose a detailed modeling approach for large \acrshort{5GDHC} networks, including pressure drops and heat losses.
However, the heat exchange between pipes and soil is not considered directly but must be provided by appropriate models using \acrfull{FMI} coupling if it is to be included.

\citet{GRIESBACH2022118696} developed a numerical model of an ice storage that supplies energy to a research building at the University of Bayreuth, Germany, using the enthalpy method.
The model is implemented in MATLAB Simulink and validated against the exact solution of the 2-phase Stefan problem and compared to measurement data of about one year \cite{GRIESBACH2022118696}.
\citet{Allan2022} presented a model for an ice storage where the available energy is calculated based on the measurement of three temperatures and the fraction of ice indirectly measured based on the height of the fluid in the storage. 
The model is used for the calculation of heat transfer coefficients of two ice thermal storages.
\citet{Carbonell2016} presented an ice storage model for solar heating applications.
The model includes a ground model, considering radial and axial discretization around the ice storage, validating the model using a pilot plant providing energy to a kindergarten.
The ice storage in this work is equipped with flat plate heat exchangers mounted vertically at the bottom of the ice storage.
Other works make use of predefined models using commercial software like TRNSYS \cite{Saini2023} or open-source tools like Modelica Buildings Library \cite{Wetter2014} and AixLib Building Systems \cite{Rogers2019} to simulate \acrshort{5GDHC} networks.
\citet{Sommer2020}, for example, simulate a reservoir network using Modelica Building Libraries model, neglecting the heat exchange and storage effect of the ground.
The consideration of the ground is a central component in the modeling of heating and cooling networks with uninsulated pipes.
However, the ground is usually neglected or only considered by a boundary condition of the unaffected ground temperature \cite{Bilardo2021}.

\subsection{Contribution of the work}
In this paper, a modeling approach is proposed to capture the dynamic effects of the interaction of \acrshort{5GDHC} network components with the surrounding soil.
The models are formulated in a software-independent way and expressed by physical principles of the components using energy and mass balances.
This allows the implementation of the model in numerous simulation and optimization environments that handle dynamic systems like Modelica.
The developed model is validated comparing simulation results with measurement data from the network in Gutach-Bleibach, Germany, which supplies heat and cold to modern buildings.
The model places a strong emphasis on the relevant dynamics of low-temperature heating networks, which may not be present in conventional heating networks. 
For example, the entire network model is equipped with a detailed ground model to map the heat exchange between the uninsulated pipes and the surrounding ground. 
This allows, for example, the use of the ground as a heat storage to be considered.
In addition, an ice storage is modeled as a central seasonal heat storage that supplies heat during the heating period and can be used to cool the buildings in summer. 
The model created for simulation purposes is compared with real measured values over a period of one year, starting in the middle of the heating period in December, and good agreement is demonstrated.
The measured data are fluid temperatures and mass flows within the network and water temperatures of the ice storage.

\subsection{Structure of the paper}
The paper is organized as follows.
First, the considered \acrshort{5GDHC} network in Gutach-Bleibach located in the south of Germany is introduced in Section~\ref{secSystem}, which contains typical components of a \acrshort{5GDHC} network and serves as inspiration for the model and is used for validation.
The network supplies heat and cold to more than 30 modern buildings using renewable energy and a central seasonal storage as well as decentral \acrshort{hps}, \acrfull{sas} and heat storage on the building site.
In Section~\ref{secModel}, the developed model is introduced.
The ice storage, used as seasonal \acrfull{ltes} and the ground are modeled in detail to account for heat exchange between the storage and pipes with the surrounding soil, which can be considered as freely available thermal storage.
Then, in Section~\ref{secResults}, the model is used to simulate the behavior of the existing network in Gutach-Bleibach for about one year of operation.
The simulation results are evaluated using temperature measurement data from the network on the basis of quality criteria, before in Section~\ref{secDiscussion} extensions and improvements are proposed and the suitability of the model for optimization tasks is discussed.

\section{System description} \label{secSystem}
In the following, the design and components of the district heating network in Gutach-Bleibach is introduced.

\subsection{\acrshort{5GDHC} network}
\begin{figure}[t!]
	\centering
	\includegraphics[width=0.9\textwidth]{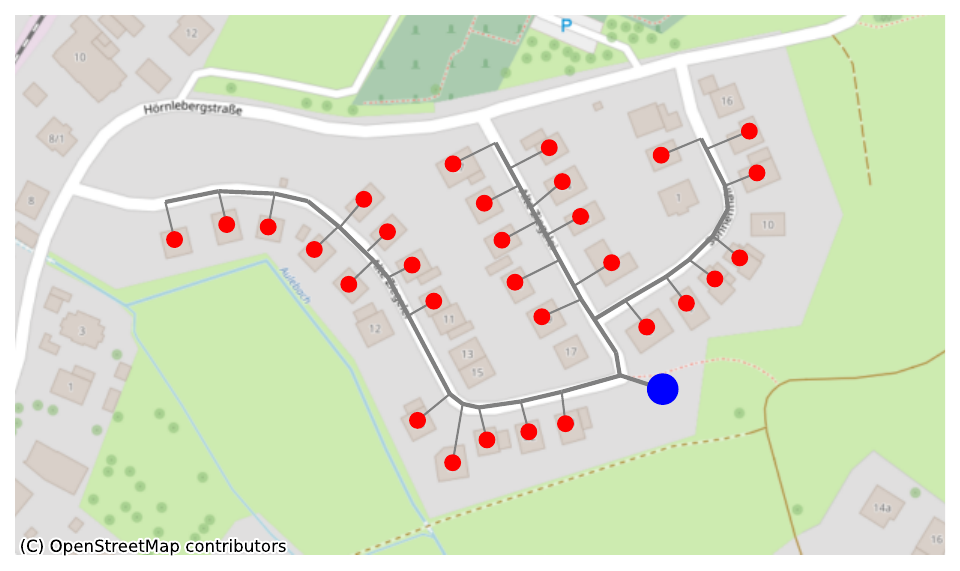}
	\caption{Illustration of the \acrshort{5GDHC} network in Gutach-Bleibach, Germany. The ice storage (blue circle) is operated as a seasonal storage to provide heat during the heating period and cooling during the summer to 30 buildings (red circles). Map based on OpenStreetMap data © OpenStreetMap contributors, modified by the authors. Used under the Open Database License \href{https://www.openstreetmap.org/copyright}{https://www.openstreetmap.org/copyright}}.
	\label{fig:Gutach}
\end{figure}

The system can be classified as a \acrshort{5GDHC} network and is located in the south of Germany.
The network consists of more than 30 recently built single- or multi-family houses in a star/radial network structure, also called a tree network \citep{Sulzer2021}.
All buildings are connected to a \acrshort{ltes}, which is an ice storage buried in the ground (see Fig.~\ref{fig:Gutach}).

The buildings are mostly equipped with \acrshort{sas} in combination with decentral short-term thermal energy storage and a brine-water \acrfull{hp}.
The buildings are hydraulically separated from the heating network by network transfer stations with heat exchangers.
The network is operated in a two-pipe configuration with a directed mass flow and a bidirectional energy flow.
The connected buildings extract heat from the grid in the case of heating and release heat into the grid in the case of cooling or in the regeneration phase.
In the heating case, the supply pipe corresponds to that of the warmer pipe compared to the return pipe.
In the cooling case or the regeneration phase, the allocation and flow direction remains the same, even if the supply line has lower temperatures than the return line.
During the regeneration phase, surplus heat from cooling or solar thermal yields is used to regenerate the ice storage.
The heat may also be stored in the ground and used to supply the buildings during the heating period.
Due to the use of crystallization heat, the system has advantages over individual heat pumps with air as a heat source, especially on very cold days with temperatures below freezing point, and also no noise nuisance due to the lack of an outdoor coil for the heat pumps.

\subsection{Components}
\begin{figure}[t!]
	\centering
	\includegraphics[width=0.6\textwidth]{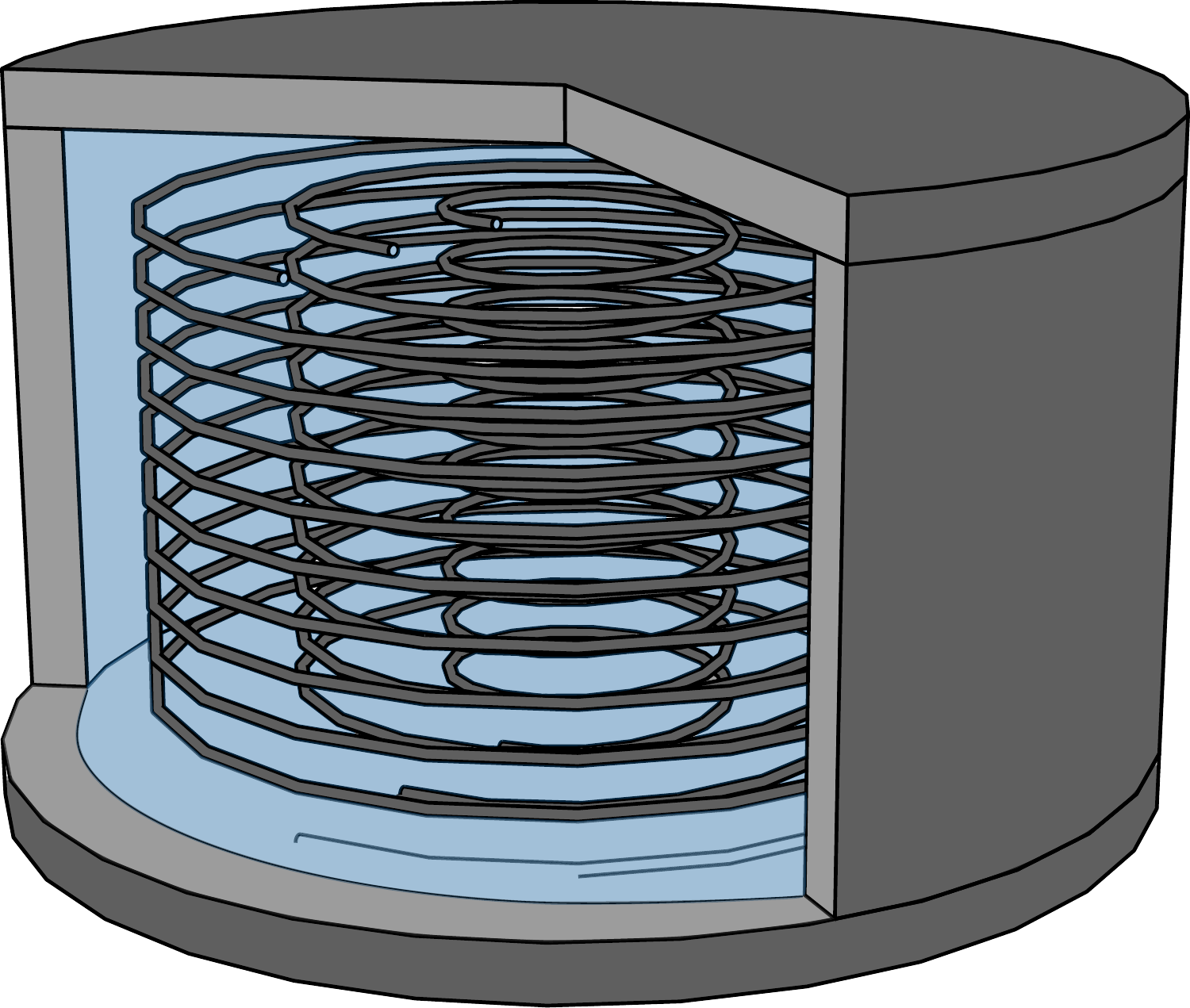}
	\caption{The ice storage is filled with water and exchanges heat with the brine flowing through the pipes. If the temperatures inside the storage reach \SI{0}{\degreeCelsius} the storage eventually starts to freeze leveraging the effect of crystallization heat. During summer the frozen storage is used to cool buildings. For the sake of simplicity, only one heat exchanger coil is shown.}
	\label{fig:is}
\end{figure}
The main components of the network considered in this work are the ice storage as a central \acrshort{ltes}, the uninsulated network pipes and the surrounding ground, as well as the transfer stations on the building site.
An ice storage is usually a water-filled cistern that is completely buried below the surface of the ground.
The walls of the storage are made of concrete and are not further insulated.
Inside the storage are large coiled pipes in which a frost-proof brine circulates as the network medium, functioning as heat exchanger between the network fluid and the water inside the storage. 
Typically, two strings of heat exchangers are used, one for discharging and one for regenerating the storage, as in \citep{Allan2022}.

\begin{figure}[b!]
	\centering
	\includegraphics[width=0.7\textwidth]{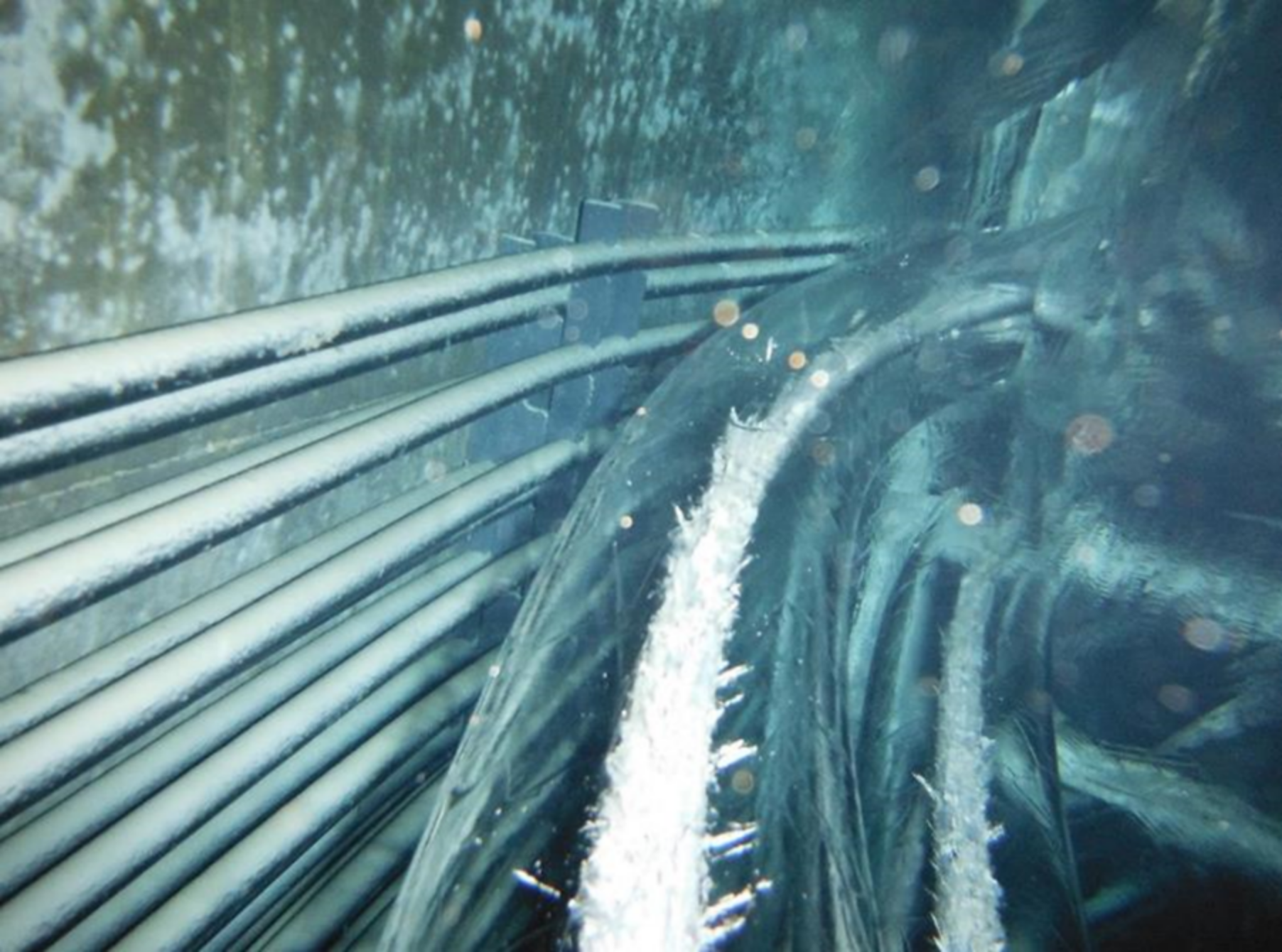}
	\caption{View of the inside of the ice store at Gutach-Bleibach during the heating season. Ice forms around the extraction heat exchanger.}
	\label{fig:isFrozen}
\end{figure}

Ice storage makes use of the heat of crystallization.
The heat that can be extracted from freezing a certain amount of water corresponds approximately to the amount of heat that can be extracted from cooling the same amount of liquid water from \SI{80}{\degreeCelsius} to \SI{0}{\degreeCelsius}.
In order to prevent cracking or even bursting of walls, the crystallization process should take place from the bottom up and from the center to the walls \cite{Allan2022}.
For this reason, the extraction heat exchanger is located in the center of the storage to allow the water to freeze from the inside to the outside and from the bottom to the top (see Fig.~\ref{fig:is} and Fig.~\ref{fig:isFrozen}), thus preventing damage to the tank structure.
The regeneration heat exchanger is positioned on the inner wall of the storage shell, where the water usually freezes later due to the greater distance between the extraction heat exchanger and the storage shell.
Heat is then transferred to the storage, thawing the ice evenly from top to bottom and from outside to inside.

The ice storage is connected to the network, supplying heat to the buildings via uninsulated plastic pipes. 
In the transfer stations, heat energy is exchanged between the network and the connected actors via heat exchangers. 
Depending on the temperature levels and operation mode, the network serves as a heat source or heat sink.

\subsection{Control Strategy}\label{subsec:CS}
The network is operated using standard rule-based control actions.
Temperature\hyp{}controlled rules are applied by the network operator.
Since the operator cannot interfere with the household energy systems, the main goal for the operation strategy is to operate the network at certain seasonal temperatures.
The temperature of the heat transfer medium may be influenced by the activation and deactivation of the central storage.
\begin{figure}[t!]
	\centering
	\includegraphics{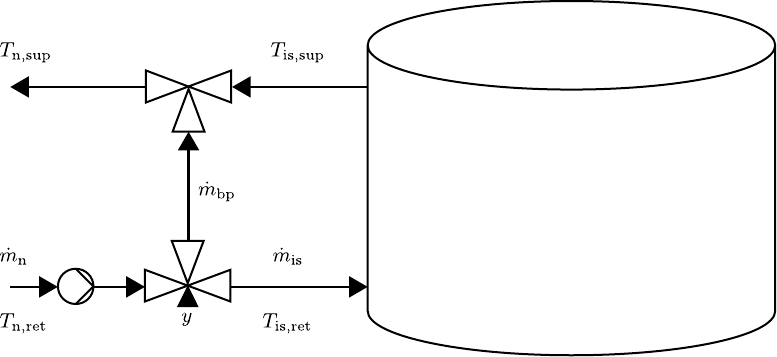}
	\caption{Schematic depiction of the pipe routing at the ice storage. The position $ y $ of the mixing valve is manipulated using a PI controller with seasonal reference temperatures for the supply temperature of the network $ T_{\n,\sup} $.}
	\label{fig:is_bp}
\end{figure}
The control variable is the position $ y $ of a mixing valve ahead of the ice storage in order to control the (proportional) activation of the storage, see Fig.~\ref{fig:is_bp}.
Depending on the position, the network mass flow $ \dot{m}_\n $ is divided into a mass flow through the ice storage $ \dot{m}_\is $ and a mass flow bypassing the storage $\dot{m}_\bp $.
At the beginning of and during the heating period, the storage is used to provide the highest possible supply temperatures.
By withdrawing energy during the heating period and the resulting reduction in temperature, the storage eventually starts to freeze.
The energy required for the regeneration of the frozen storage can be supplied in the form of cooling demands from the buildings during the summer.
The activation and deactivation of the thermal storage is therefore controlled by seasonal temperature specifications (set-point) for the network supply temperature $ T_{\n,\sup} $ (controlled variable) using a PI controller for the change of the valve position $ y $ (manipulated variable) based on the error, as well as manual interaction.

\section{Model description}\label{secModel}
This section introduces the developed models of the \acrshort{5GDHC} network components.
The models are based on mass and energy balances, formulated as \acrfull{odes}.
This includes models for the pipe network, the ice storage, the surrounding soil and heat transfer stations.
The models are spatially discretized, whereby the system is divided into discrete sections or layers in order to capture the spatial variation of the variables under consideration.

\subsection{Pipe model}
\label{subsec:pipe}
A directed mass flow is assumed for the pipe network of the heating network models.
The pipe model is described using a single pipe, but applies to both supply and return pipes unless otherwise noted.

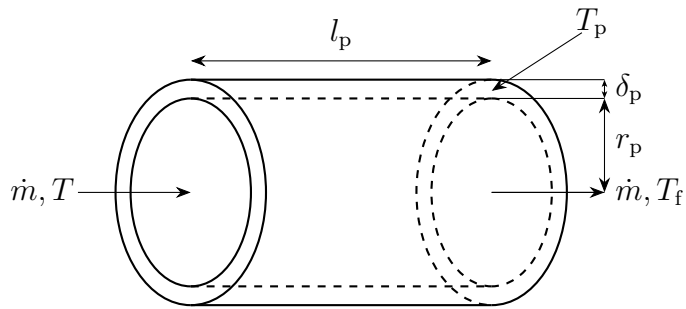
\begin{figure}[b]
	\centering
	\begin{tikzpicture}

		\def\thicknessT{0.25}     
		\def\radiusT{1.25}        
		\def\lengthT{4}          
		\def\totalRadius{\radiusT + \thicknessT}
		\def\totalNegRadius{\radiusT - \thicknessT}

		\draw [thick](-\lengthT/2,+\totalRadius) -- (+\lengthT/2,+\totalRadius);
		\draw [thick](-\lengthT/2,-\totalNegRadius) -- (+\lengthT/2,-\totalNegRadius);

		\draw [thick](-\lengthT/2,+\totalRadius) arc (90:270:1 and \totalRadius);
		\draw [thick](-\lengthT/2,+\totalRadius) arc (90:-90:1 and \totalRadius);

		\draw [thick, dashed](+\lengthT/2,+\totalRadius) arc (90:270:1 and \totalRadius);
		\draw [thick](+\lengthT/2,+\totalRadius) arc (90:-90:1 and \totalRadius);

		\draw [thick, dashed](-\lengthT/2,+\radiusT) -- (+\lengthT/2,+\radiusT);
		\draw [thick, dashed](-\lengthT/2,-\radiusT) -- (+\lengthT/2,-\radiusT);

		\draw [thick](-\lengthT/2,+\radiusT) arc (90:270:0.8 and \radiusT);
		\draw [thick](-\lengthT/2,+\radiusT) arc (90:-90:0.8 and \radiusT);

		\draw [thick, dashed](+\lengthT/2,+\radiusT) arc (90:270:0.8 and \radiusT);
		\draw [thick, dashed](+\lengthT/2,+\radiusT) arc (90:-90:0.8 and \radiusT);

		\draw [thin] (+\lengthT/2,+\totalRadius) -- (+\lengthT/2+\totalRadius, \totalRadius);
		\draw [thin] (+\lengthT/2,+\radiusT) -- (+\lengthT/2+\totalRadius, \radiusT);
		\draw [thin] (+\lengthT/2,0) -- (+\lengthT/2+\totalRadius,0);

		\draw [{Stealth[length=2mm]}-{Stealth[length=2mm]}] (-\lengthT/2,+\totalRadius + 0.25) -- node[midway, above] {$l_\tu$} (\lengthT/2, \totalRadius + 0.25);

		\draw [{Stealth[length=2mm]}-{Stealth[length=2mm]}] (+\lengthT/2+\totalRadius,0) -- node[midway, right]{$r_\tu$} (+\lengthT/2+\totalRadius,\radiusT);
		\draw [{Stealth[length=1mm]}-{Stealth[length=1mm]}] (+\lengthT/2+\totalRadius,\radiusT) -- node[midway, right] {$\delta_\tu$} (+\lengthT/2+\totalRadius,\totalRadius);
		
		\draw [-{Stealth[length=2mm]}] (+\lengthT/2,0) --(+\lengthT/2+\totalRadius,0);
		\node at (+\lengthT/1.4+\totalNegRadius+\thicknessT,0) {$\dot{m},T_\f$};

		\draw [-{Stealth[length=2mm]}] (-\lengthT/2-\totalNegRadius,0) -- (-\lengthT/2,0);
		\node at (-\lengthT/1.8-\totalNegRadius-\thicknessT,0) {$\dot{m},T$};
		
		\draw [-{Stealth[length=2mm]}] (\lengthT-0.8,1.9*\totalNegRadius) --(\lengthT-2,\totalNegRadius+\thicknessT+.1);
		\node at (+\lengthT/1.2,2*\totalNegRadius) {$T_\tu$};

	\end{tikzpicture}
	\caption{Sketch of the pipe geometry.}
	\label{fig:pipe}
\end{figure}

The geometric properties of a pipe are calculated using formulas for cylindrical and hollow cylindrical objects (see Fig.~\ref{fig:pipe}).
The cross section area $ A_\f $ and volume $ V_\f $ of the inner pipe is calculated according to equations \eqref{eq:A_f} and \eqref{eq:V_f} using the radius $r_\tu$ and length $l_\tu $ of the pipe.
The mass $ m_\f $ of the fluid within a pipe is calculated by \eqref{eq:m_f}, with $ \rho_\f $ being the density of the fluid.

The surface area $ A_\tu $ and volume $ V_\tu $  of the pipe are calculated using the respective formulas for hollow cylinders \eqref{eq:A_t} and \eqref{eq:V_t}, where $ \delta_\tu $ is the thickness of the pipe.
The mass of the pipe $ m_\tu $ is then described by the density of the pipe material $ \rho_{\tu} $ and the volume of the pipe $ V_{\tu} $ \eqref{eq:m_t}.
\begin{subequations}
\begin{align}
	 & A_\f = \pi r_\tu^2 \label{eq:A_f} \\
	 & V_\f =  A_\f l_\tu \label{eq:V_f} \\
	 & m_\f = \rho_\f V_\f \label{eq:m_f} \\
	 & A_\tu = 2\pi r_\tu l_\tu \label{eq:A_t} \\
	 & V_\tu = \pi \bigl((r_\tu + \delta_\tu)^2 - r_\tu^2\bigr) l_\tu \label{eq:V_t} \\
	 & m_\tu = \rho_\tu V_\tu \label{eq:m_t}
\end{align}
\end{subequations}

The temperature-dependent heat transfer in the heating network's pipe system is modeled using the Method of Lines \cite{Budak1962}.
This involves discretizing the pipe system spatially into finite segments and then carrying out a finite volume formulation.
The resulting energy equations form an \acrfull{ode} system.
This approach enables the dynamic heat propagation throughout the entire network to be described efficiently and modularly, as demonstrated for example in \cite{Betancourt2019}.
The temporal integration can be carried out in simulation using suitable solvers.

\subsubsection{Fluid}
The temperature development of the fluid running through the pipe $T_\f$ is determined by
\begin{subequations}
\begin{align}
	\frac{\dn T_\f(t)}{\dt} & = \frac{\dot{m}(t) c_\f \Bigl(T(t) - T_\f(t)\Bigr) - \dot{Q}_{\f,\tu}(t)}{m_\f c_\f}, \label{eq:T_f} \\
	\dot{Q}_{\f,\tu}(t)     & = \frac{2 \pi l_\tu}{\frac{1}{r_\tu\alpha_\tu} + \frac{1}{\lambda_{\tu}} \ln \Bigl(\frac{r_{\tu} + \delta_{\tu}/2}{r_{\tu}}\Bigr)} \Bigl(T_\f(t) - T_\tu(t) \Bigr). \label{eq:Qdot_ft}
\end{align}
\end{subequations}
Herein, $ c_\f $ is the specific heat capacity of the network fluid.
The state $ T $ corresponds to the temperature of the medium entering the pipe volume.
The mass flow of the medium entering and leaving the pipe is described by $ \dot{m} $.
The temperature is also influenced by the heat flow between the fluid and the pipe $ \dot{Q}_{\f,\tu}$, with $ T_\tu $ being the temperature of the pipe.
The temperature of a finite volume such as a pipe is calculated at its center.
Accordingly, when calculating the temperature of the fluid, a heat convection component between the temperature of the medium and the temperature of the pipe and a heat conduction to the center of the pipe must be taken into account.
The coefficient of heat transfer for the heat convection part is denoted by $\alpha_{\tu} $.
The heat conduction is influenced by the thermal conductivity of the pipe $ \lambda_{\tu} $.

\subsubsection{Pipe}
For modeling the energy balance of the pipe, the heat flow between the pipe and soil $\dot{Q}_{\tu,\s}$ has to be considered as well.
Therefore, the pipe temperature $ T_{\tu} $ is described by
\begin{equation}
	\frac{\dn T_{\tu}(t)}{\dt} = \frac{\dot{Q}_{\f,\tu}(t) - \dot{Q}_{\tu,\s}(t)}{m_{\tu} c_{\tu}}, \label{eq:T_t}
\end{equation}
with $ c_\tu $ being the specific heat capacity of the pipe material.
The heat exchange $ \dot{Q}_{\tu,\s} $ with the surrounding soil takes place via heat conduction:
\begin{align}
	\begin{split}
	\dot{Q}_{\tu,\s}(t) = &\frac{2 \pi l_\tu}{\frac{1}{\lambda_{\tu}} \ln \Bigl(\frac{r_{\tu} + \delta_{\tu}}{r_{\tu}+ \delta_{\tu}/2} \Bigr) + \frac{1}{\lambda_{\s}} \ln \Bigl(\frac{r_{\tu} + \delta_{\tu} + \delta_{\s}/2}{r_{\tu}+ \delta_{\tu}}\Bigr)} \\ 
	& \cdot \Bigl(T_\tu(t) - T_{\s,1}(t) \Bigr), \label{eq:Qdot_ts}
	\end{split}
\end{align}
where $ \delta_{\s} $ is the thickness of one soil layer, $ \lambda_{\s} $ the thermal conductivity of soil and $ T_{\s,1} $ the temperature of the soil for the first layer surrounding the pipes.

\subsubsection{Pressure loss calculation}
The Darcy-Weisbach equation, in an adjusted form of \cite{Eckert2021}, is used to calculate the pressure drop $\Delta p $ in incompressible flow through the pipe
\begin{subequations}
	\begin{align}
		\Delta p(t) &= \frac{8 l_\tu {\dot{m}(t)}^2}{\rho_\f \pi^2 (2 r_\tu)^5} f(t), \label{eq:deltaP} \\
		f(t) &= 0.3164 \Rey(t)^{-\frac{1}{4}}, \label{eq:DarcyFriction} \\
		\Rey(t) &= \frac{\dot{m}(t) 2 r_\tu}{\mu_\f(t) A_\f} \label{eq:Rey},
	\end{align}
\end{subequations}
with $ f $ being the Darcy friction factor \eqref{eq:deltaP}.
The Darcy friction factor \eqref{eq:DarcyFriction} is calculated according to the Blasius correlation \cite{Hager2003}.
The Reynolds number Re is calculated according to \eqref{eq:Rey} adapted from \cite{Kuchling2022}, making the Reynolds number dependent on the mass flow $ \dot{m} $, the diameter of the pipe, the dynamic viscosity $ \mu_\f $ and the cross-sectional area of the pipe $ A_\f $.
Since the temperature of the fluid in low-temperature networks may drop below the freezing point of water, a water-glycol mixture with a lower freezing point is generally used. 
Depending on the mixture ratio, the viscosity is significantly higher than that of water. 
The dynamic viscosity of the fluid is also highly temperature-dependent and can be calculated by interpolating data points from the fluid data sheet.

\subsection{Ground model}\label{subsec:ground}
The soil surrounding the pipe is a key component when uninsulated pipes are used in \acrshort{5GDHC} networks because of the thermal exchange between the fluid in the pipes and the soil.
The soil and ground as a whole may be used as additional thermal storage.
Due to the low network temperature, uninsulated pipes are used to enhance heat exchange between the pipe and the surrounding soil.
This enables excess heat to be stored during the regeneration phase and used during heating periods.
In order to depict this behavior, a comprehensive model of the ground is essential, especially if the heat source of the network is mainly based on shallow geothermal energy through the network pipes.

\begin{figure}
	\centering
	\includegraphics[width=0.8\textwidth]{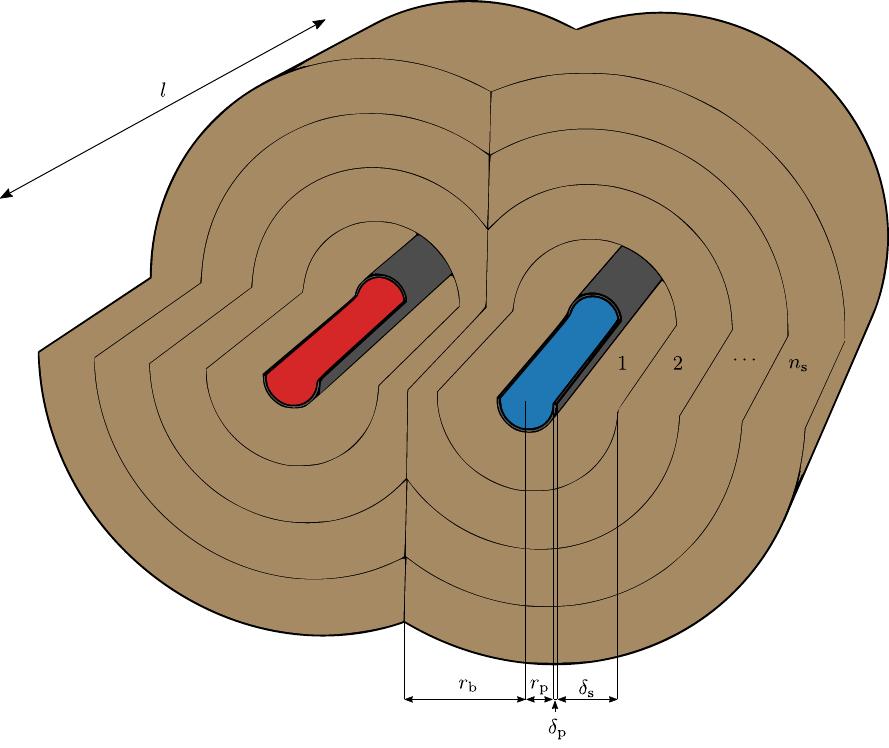}
	\caption{Graphical representation of the modeled soil surrounding the network pipes. The soil is radially discretized into $ n_\s $ cylindrical shapes, taking into account the intersecting layers between the supply (red medium) and return (blue medium) sides.}
	\label{fig:soil}
\end{figure}

Within this work, the ground surrounding the pipes is modeled using energy balances, explicitly accounting for the heat distribution perpendicular and parallel to the pipe network as well as the interacting behavior of the supply and return sides of the ground influenced by the fluid and pipe temperatures (cf. Section~\ref{subsec:pipe}).
A graphical representation of the modeled ground surrounding a pipe section is shown in Fig.~\ref{fig:soil}.
The model considers the interaction of the supply and return pipe sides with the surrounding soil, as well as the interaction between the supply and return sides.

The soil surrounding the supply and return sides is further divided into two sections.
One part of the soil represents the respective outer soil layers in relation to the intersection between the supply and return sides.
These parts are not adjacent to the opposite side and therefore have no direct energy exchange.
A further section adjoins each of these, which has a direct exchange between the supply and return sides.
The two sections on the supply and return sides also have a direct heat transfer (see Fig.~\ref{fig:soil2}).

\subsubsection{Geometrical properties}

\begin{figure}
	\centering
	\includegraphics{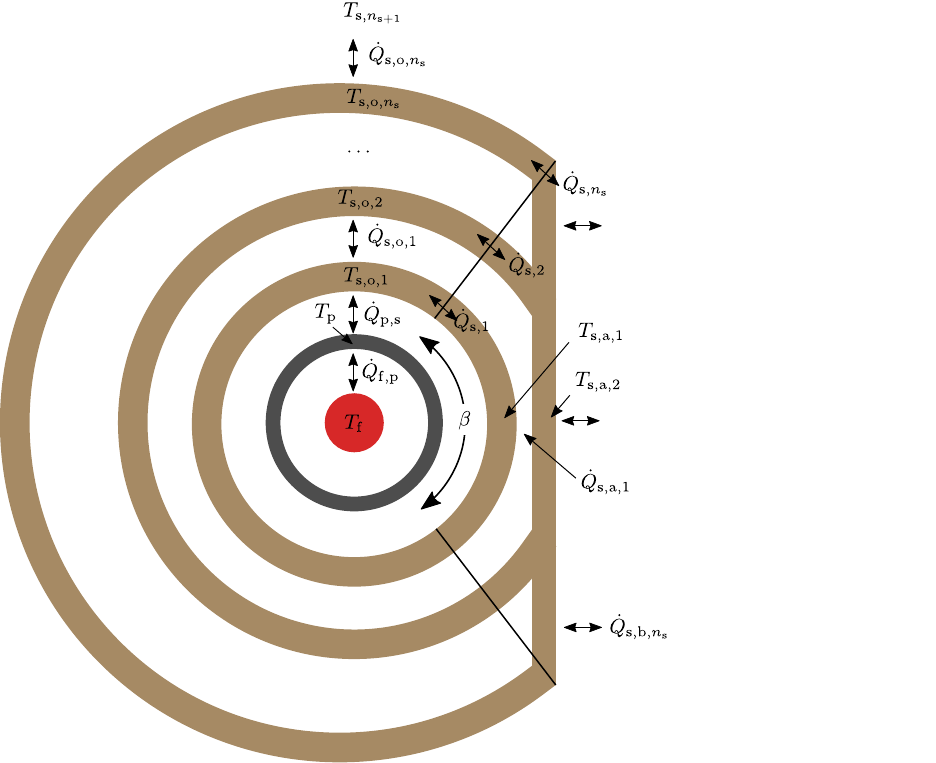}
	\caption{Schematic representation of the soil volumes surrounding the supply line. The soil is discretized into $ n_{\s} $ layers and further divided into an outer part that is not directly influenced by the return side (o) and an adjacent part that is directly influenced by it (a).}
	\label{fig:soil2}
\end{figure}

The soil around the pipes is discretized perpendicular to the pipes into $n_\mathrm{s}$ layers.
If the considered perpendicular length of a discretized soil layer around the pipes is less than the installation distance $ r_\mathrm{b} $  between them, the hollow cylindrical shapes of the soil layers are reduced by the intersecting circular segments.
This is only considered for the adjacent sections (see Fig.~\ref{fig:soil2}) and allows for possible interactions and effects of temperature gradients between the supply and return sides of the ground.
The geometrical properties of the soil layers are calculated by
\begin{subequations}
	\begin{align}
		r_{\s,i}  & = r_{\tu} + \delta_{\tu} + i \delta_{\s},                         & i = 0,\dots, n_{\s}, \label{eq:r_s} \\
		z_{\s,i}  & = \mathrm{max}(r_{\s,i} - r_\mathrm{b}, 0),                   & i = 0,\dots, n_{\s}, \label{eq:h}   \\
		l_{\mathrm{ch},\s,i} & = 2 \sqrt{2 r_{\s,i} z_{\s,i} - z_{\s,i}^2},                     & i = 0,\dots, n_{\s}, \label{eq:cl}  \\
		s_i       & = 2 r_{\s,i} \arcsin \biggl(\frac{l_{\mathrm{ch},\s,i}}{2 r_{\s,i}}\biggr), & i = 0,\dots, n_{\s}. \label{eq:s_k}
	\end{align}
\end{subequations}

The radius $ r_{\s,i} $ of soil layer $ i $ is composed of the radius $ r_\tu $ and thickness $ \delta_\tu $ of the pipe, as well as a corresponding number of thickness $ \delta_{\s} $ of the soil layers \eqref{eq:r_s}.
If the radius $ r_{\s,i} $ exceeds the installation radius $ r_\mathrm{b} $ between the supply and return pipe, the respective layer $ i $ of the supply and return side intersects each other.
The hollow cylindrical shapes of the ground layers are then subtracted by the circular segments accounting for the intersecting ground area.
Therefore, the height $ z_{\s} $ of the circular has to be determined \eqref{eq:h}.
The height is then used to calculate the chord length, $ l_{\mathrm{ch},\s,i} $ \eqref{eq:cl}.
Finally, the arc length $ s_{i} $ is defined as \eqref{eq:s_k}.

The endpoints of the chord define the boundary between the outer and adjacent soil regions. Geometrically, the outer soil regions correspond to the remaining area of the circle after subtracting the circular sectors associated with the adjacent regions, whereas the adjacent regions are reduced by the circular segments formed by the intersection of the supply and return sides (see Fig.~\ref{fig:soil2}).
Hence, the angle for the circular segment is determined by
\begin{align}
	\beta = 2 \arcsin\Bigl(\frac{l_{\mathrm{ch},\s,n_\s}}{2 r_{\s,n_\s}}\Bigr). \label{eq:beta}
\end{align}

Depending on the respective radius of the soil layers, the surface area of a discretized soil layer $ A_{\mathrm{s},\{\adj,\out\},i} $ for $ i=1,\dots,n_{\s} $ is determined by
\begin{subequations}
	\begin{align}
		A_{\mathrm{s},\out,i} = &  \pi (r_{\s,i}^2 - r_{\s,i-1}^2) \frac{360\degree - \beta}{360\degree}, \label{eq:A_s_1} \\
		A_{\mathrm{s},\adj,i} = & \left( \pi r_{\s,i}^2 \frac{\beta}{360\degree} - \Big( r_{\s,i}^2 \arcsin\Bigg( \frac{l_{\mathrm{ch},\s,i}}{2r_{\s,i}} \Bigg) - \frac{l_{\mathrm{ch},\s,i}(r_{\s,i} - z_{\s,i})}{2} \Big) \right) \nonumber \\                                                   
		& -  \Bigg( \pi r_{\s,i-1}^2 \frac{\beta}{360\degree}    \\
		& \hspace{0.4cm} - \Big( r_{\s,i-1}^2 \arcsin \Bigg( \frac{l_{\mathrm{ch},\s,i-1}}{2r_{\s,i-1}} \Bigg) - \frac{l_{\mathrm{ch},\s,i-1}(r_{\s,i-1} - z_{\s,i-1})}{2} \Big) \Bigg). \nonumber \label{eq:A_s_2} 
	\end{align}
\end{subequations}

The area for the outer section $ A_{\s,\out,i} $ of the soil equals the circular sector depending on the circular segment angle $ \beta $ \eqref{eq:beta}.
For the area of the adjacent section $ A_{\s,\adj,i} $, the circular segment is subtracted according to the overlap of the supply and return side.

In order to account for the reduced area and volume of the respective layers relevant to the heat flow, a correction factor $ k_{\{\adj,\out\},i} $, proportional to the area of an unchanged hollow cylinder $ A_{\h,i} $, is introduced:
\begin{subequations}
\begin{align}
	 k_{\{\adj,\out\},i} &= \frac{A_{\s,\{\adj,\out\},i}}{A_{\h,i}}, & i = 1,\dots, n_{\s}, \\
	 A_{{\h},i} &= \pi (r_{\s,i}^2 - r_{\s,i-1}^2), & i = 1,\dots, n_{\s}, \label{eq:A_hollow} \\
	 V_{\s,\{\adj,\out\},i} &= A_{\s,\{\adj,\out\},i} l_\tu, & i=1,\dots,n_{\s},
\end{align}
\end{subequations}
with $ k_{\{\adj,\out\},i} \in [0,1] $ and $ V_{\s,\{\adj,\out\},i}  $ being the volume of the respective soil layers.

\subsubsection{Soil}

For formulating the energy balances of the soil layers, the temperatures of the soil are determined by
\begin{subequations}
\begin{align}
	\frac{\dn T_{\s,\out,1}(t)}{\dt} = &\frac{\dot{Q}_{\tu,\s}(t)k_{\out,1} - \dot{Q}_{\s,\out,1}(t) - \dot{Q}_{\s,1}(t)}{V_{\s,\out,1} C_{\s,\out,1}(T_{\s,\out,1})},
	 \label{eq:T_s11}  \\                 
	\frac{\dn T_{\s,\adj,1}(t)}{\dt} = & \frac{\dot{Q}_{\tu,\s}(t)k_{\adj,1} - \dot{Q}_{\s,\adj,1}(t) + \dot{Q}_{\s,1}(t) - \dot{Q}_{\s,\mathrm{b},i}}{V_{\s,\adj,1} C_{\s,\adj,1}(T_{\s,\adj,1})}, \label{eq:T_s12} \\
	\frac{\dn T_{\s,\out,i}(t)}{\dt} = & \frac{\dot{Q}_{\s,\out,i-1}(t) - \dot{Q}_{\s,\out,i}(t) - \dot{Q}_{\s,i}(t)}{V_{\s,\out,i} C_{\s,\out,i}(T_{\s,\out,i})}, \label{eq:T_s1} \\
	& i=2,\dots,n_{\s}, \nonumber \\
	\frac{\dn T_{\s,\adj,i}(t)}{\dt}         = & \frac{\dot{Q}_{\s,\adj,i-1}(t) - \dot{Q}_{\s,\adj,i}(t) + \dot{Q}_{\s,i}(t) - \dot{Q}_{\s,\mathrm{b},i}}{V_{\s,\adj,i} C_{\s,\adj,i}(T_{\s,\adj,i})}, \label{eq:T_s2} \\
	& i=2,\dots,n_{\s}-1, \nonumber \\
	\frac{\dn T_{\s,\adj,n_{\s}}(t)}{\dt}         = & \frac{\dot{Q}_{\s,\adj,n_{\s}-1}(t) + \dot{Q}_{\s,n_{\s}}(t) - \dot{Q}_{\s,\mathrm{b},n_{\s}}}{V_{\s,\adj,n_{\s}} C_{\s,\adj,n_{\s}}(T_{\s,\adj,n_{\s}})}. \label{eq:T_s2_ns} 
\end{align}
\end{subequations}
The temperatures of the first soil layers $ T_{\s,\{\adj,\out\},1} $ are influenced by the heat exchange with the pipe $ \dot{Q}_{\tu,\s} $, the heat exchange between the first and second layers within each section $ \dot{Q}_{\s,\{\adj,\out\},1} $ and the exchange between the outer and adjacent section $ \dot{Q}_{\s,1} $ \eqref{eq:T_s11} - \eqref{eq:T_s12}.
The temperatures of the adjacent soil layers of the supply and return side further have a direct heat flow $ \dot{Q}_{\s,\mathrm{b}} $ \eqref{eq:T_s12}.
The temperatures for the other layers of soil are defined for the outer parts according to \eqref{eq:T_s1} and for the adjacent parts according to \eqref{eq:T_s2}.
An uninfluenced soil temperature $ T_{\s,\ns+1} $ is assumed at the boundary of the soil, that is only relevant for the outer section as a boundary condition \eqref{eq:T_s2_ns}.
The calculation of the volumetric heat capacity $ C_{\mathrm{s},\{\adj,\out\},i}(T_{\mathrm{s},\{\adj,\out\},i}) $ is adapted from \cite{BodenEnergiewende2015}, \cite{Sres2009} and calculated by
\begin{align}
	\label{eq:C_s}
	C_{\mathrm{s},\{\adj,\out\},i}(T_{\mathrm{s},\{\adj,\out\},i}) &= \\
	&\begin{cases} 
		\rho_{\s} \bigl( (1-w) c_{\s,\dr} + c_{\w} w \bigr),               \text{if } T_{\s,\{\adj,\out\},i}(t) > T_{\w,\mathrm{fluid}}, \\
		\rho_{\s} \bigl( (1-w) c_{\s,\dr} + c_{\ice} w \bigr),         \text{if } T_{\s,\{\adj,\out\},i}(t) < T_{\w,\mathrm{solid}},  \\
		\rho_{\s} \bigl( (1-w) c_{\s,\dr} + \Delta h_{\fus}w \bigr),  \text{otherwise,} \nonumber
	\end{cases} 
\end{align}
for $ i = 1,\dots,n_{\s} $.
The heat capacity is derived from the ratio of water $ w \in [0, 1] $ of the soil and the specific heat capacity for dry soil $ c_{\s,\mathrm{d}} $.
Different heat capacities are applied depending on the temperature and aggregate state of the water in the ground.
It is assumed that the water freezes at a temperature of $ T_{\w,\mathrm{fluid}} = \SI{0}{\degreeCelsius} $ and is completely frozen at $ T_{\w,\mathrm{solid}} = \SI{-1}{\degreeCelsius} $.
For completely liquid water, the heat capacity of water is $ c_{\w} $, whereas for frozen water it is that of ice $ c_{\ice} $.
During the freezing process, the specific enthalpy of fusion normalized to a temperature range of \SI{1}{\kelvin}, $ \Delta h_{\fus} $ is applied.
The change in volume and expansion of the water during the freezing process are neglected.

In order to account for the vertically propagating heat flow from soil layer 1 to $ n_\mathrm{s} $ the heat flows between the respective adjacent layers are determined by
\begin{align}
	\begin{split}
		\dot{Q}_{\s,\{\adj,\out\},i}(t)    = & \frac{2 \lambda_{\s} \pi l_\tu}{\ln \Bigl(\frac{r_{\tu} + \delta_{\tu} + (i+1/2) \delta_{\s}}{r_{\tu} + \delta_{\tu} + (i-1/2)\delta_{\s}} \Bigr)} \\
		& \cdot \Bigl( T_{\s,\{\adj,\out\},i}(t) - T_{\s,\{\adj,\out\},i+1}(t) \Bigr) k_{\{\adj,\out\},i,}  \label{eq:Qdot_s}
	\end{split}
\end{align}
for $ i=1, \dots, n_{\s}-1 $.

The boundary temperature  $ T_{\s,n_\s+1} $ only influences the outer section:
\begin{align}
	\begin{split}
	\dot{Q}_{\s,\adj,n_\s}(t)    = & \frac{2 \lambda_{\s} \pi l_\tu}{\ln \Bigl(\frac{r_{\tu} + \delta_{\tu} + n_\s \delta_{\s}}{r_{\tu} + \delta_{\tu} + (n_\s-1/2)\delta_{\s}} \Bigr)} \\
	&\cdot \Bigl( T_{\s,\adj,n_\s}(t) - T_{\s,\adj,n_\s+1}(t) \Bigr) k_{\adj,n_\s}.  \label{eq:Qdot_s1_ns}
	\end{split}
\end{align}

The heat flows between the respective outer and adjacent sections are calculated according to the formula for heat convection of cylindrical shapes.
Soil layers of the outer and adjacent section exchange heat via $ \dot{Q}_{\s,i} $, calculated by
\begin{align}
	\dot{Q}_{\s,i}(t) = \lambda_{\s} \frac{A_{\s}}{d_{\s,i}} \Bigl(T_{\s,\out,i}(t) - T_{\s,\adj,i}(t) \Bigr), \label{eq:Qdot_s12}
\end{align}
for $ i = 1,\dots,n_{\s} $, where $ A_{\s} = \delta_{\s} l_\tu $ is the area of a rectangular shape for the considered heat exchange and $ d_{\s,i} $ the respective distance for the heat exchange.

The heat flow term $ \dot{Q}_{\s,\mathrm{b},i} $ refers to the heat flow between the supply and return side:
\begin{align}
	\dot{Q}_{\s,\mathrm{b},i}(t) = 
	\begin{cases}
	 0, & \text{if } z_{\s,i} = 0, \\
	\lambda_{\s} \frac{A_{\s,\mathrm{b},i}}{d_{\s,\mathrm{b}}} \Bigl(T_{\s,\adj,i,\sup}(t) - T_{\s,\adj,i,\ret}(t) \Bigr), & \text{otherwise,}\label{eq:Qdot_sb}
	\end{cases}
\end{align}
for $ i=1,\dots,\ns $.
The area $ A_{\s,\mathrm{b},i} = (z_{\s,i} - z_{\s,i-1}) l_\tu $ corresponds to the intersecting area of the return and supply sides for the respective layers and $ d_{\s,\mathrm{b},i} $ the distance between the centers of the neighboring supply and return layers.
Note that for \eqref{eq:Qdot_sb} a distinction between the supply and return temperatures is required.
The equations \eqref{eq:T_s11} - \eqref{eq:Qdot_s12} apply to both the supply and return sides, given the corresponding variables for the temperatures.

\subsubsection{Boundary condition of the ground model}

\begin{figure}[t!]
	\centering
	\includegraphics[width=0.6\textwidth]{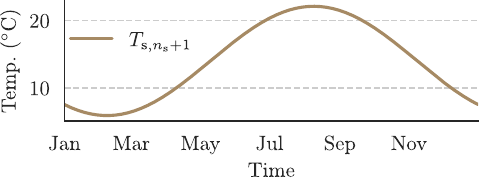}
	\caption{The soil temperature at the boundary of the considered ground is assumed to be the uninfluenced soil temperatures $ T_{\s,n_{\s+1}} $, represented by sinusoidal trend over the year. The curve corresponds to the temperature evolution calculated by \eqref{eq:T_s_bd} using the minimum (\SI{5.9}{\degreeCelsius}) and maximum (\SI{22.1}{\degreeCelsius}) soil temperature in \SI{1}{\meter} depth for the nearby Deutsche Wetterdienst site Freiburg, Germany in 2024 \cite{DWD}.}
	\label{fig:T_s_bd}
\end{figure}

The boundary of the soil is modeled with a uniform temperature $ T_{\s,n_{\s}+1} $ that applies around the entire ground model:
\begin{subequations}
\begin{align}
	T_{\s,n_{\s}+1}(t) = &T_{\s,\mathrm{min}} + a(t) ( T_{\s,\mathrm{max}} - T_{\s,\mathrm{min}}), \label{eq:T_s_bd} \\
	a(t) = &\Biggl(-\frac{1}{2} \cos \biggl( 2 \pi \Bigl( \frac{t_0 + t}{3600} - 900 \Bigr)\frac{1}{8760} \biggr) + \frac{1}{2} \Biggr).
\end{align}
\end{subequations}
The temperature is calculated using the uninfluenced soil temperature at the installation depth using measurements from weather stations nearby.
The development of $ T_{\s,n_{\s}+1} $ models the typical seasonal behavior of the shallow soil temperature representing a sinusoidal curve (Fig.~\ref{fig:T_s_bd}).
The temperatures $ T_{\s,\mathrm{\{min,max\}}} $ refer to the minimum and maximum temperature of the soil at a respective location and installation depth of the pipes.
The elapsed time $ t $ in seconds is augmented by a term $ t_0 $ for the elapsed time since the beginning of the year.

Uninsulated network pipes are also used as ground to utilize horizontal near-surface geothermal energy with an unsealed surface.
The collector output is thereby influenced by the temperature of the surrounding soil, as well as its thermophysical parameters, such as thermal conductivity and specific heat capacity.
These points are taken into account in the model and can be adapted to local conditions with adjustable parameters.
Other influencing factors are the exposure of the soil surface to sunlight, the groundwater level and energy input through precipitation.
While the first point is not currently taken into account in the model, the water content can be variably adjusted.
As most of the Gutach-Bleibach network pipes are covered by roads, the energy input from precipitation is disregarded. 
However, this is usually insignificant anyway, as will be demonstrated below.
Assuming that the temperature of the precipitation corresponds to the average outdoor temperature and taking the measurements of the ambient temperature for the precipitation and the soil temperature at a level of \SI{1}{\meter} of nearby DWD weather stations, the thermal influence of precipitation corresponds to
\begin{equation}
	\begin{split}
	Q_{\p} &= \SI{785.6}{\frac{\kilogram}{\meter^2a}} \SI{4.18}{\frac{\kilo\joule}{\kilogram\kelvin}}(\SI{285.87}{\kelvin} - \SI{287.06}{\kelvin})\\ &= \SI{-1.085}{\frac{\kilo\watt\hour}{\meter^2 a}}.
	\end{split}
\end{equation}
Assuming an average extraction capacity of geothermal collectors of $ \SI{30}{\frac{\watt}{\meter^2}} $ \cite{VDI4640} and 2000 operating hours per year, the heat input or cooling by precipitation corresponds to less than two percent, and is therefore not taken into account in the model.

\subsection{Ice storage}

\begin{figure}[t!]
	\centering
	\includegraphics[width=1\textwidth]{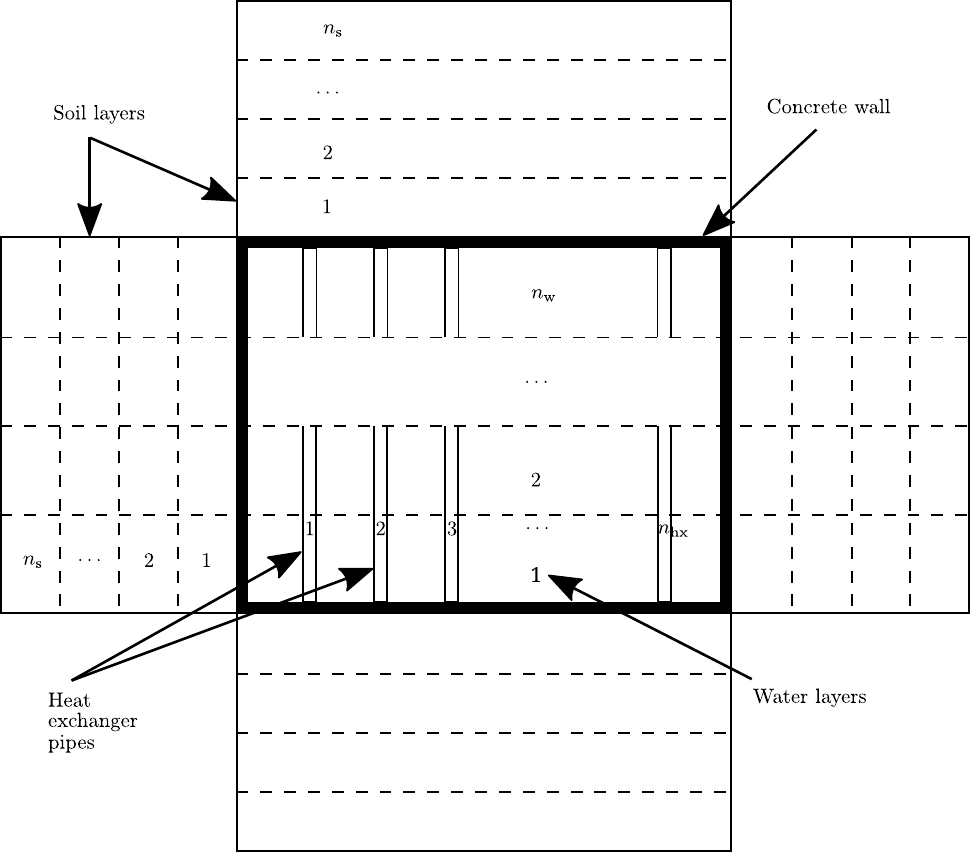}
	\caption{Schematic depiction of the discretization of the ice storage and its surrounding soil. The black rectangle with the thick edges represents the concrete wall of the ice storage.  Within the ice storage, discretization takes place along the vertical in $ \nw $ water layers. It is assumed that $ n_\hx $ exchanger coils run through the ice storage along the vertical. The soil around the ice storage is also discretized horizontally and vertically in $ n_\s $ soil layers and their interaction with the ice storage is modeled.}
	\label{fig:is_dic}
\end{figure}

The ice storage and its piping are modeled using the geometric properties of cylindrical and circular shapes.
The ice storage model is discretized along its height into $ n_{\w} $ layers to account for stratification of storage temperatures based on \cite{Eicker2003}.
The ice storage is completely embedded in the ground and is surrounded by soil on all sides. 
A schematic representation of the spatial discretization of the storage and the ground can be seen in Fig.~\ref{fig:is_dic}.

The network pipes are connected to the pipe coils in the ice storage via a distribution bar, through which the network medium flows.
The supply and return pipes that connect the network to the storage are split into $ n_\hx $ heat exchanger coils that are usually installed as multiple coiled lines inside the storage and are then recombined at the outlet.
Consequently, the total heat exchange is calculated for each water layer using $ n_\hx $ heat exchanger pipes, whose properties are calculated using one representative heat exchanger pipe per water layer in the model.
Ice storage usually includes two separate heat exchangers: one for extraction and one for regeneration.
During the extraction phase, the temperature of the water in the storage eventually drops to \SI{0}{\degreeCelsius} and ice forms around the pipes of the extraction heat exchanger, while the regeneration heat exchanger may remain ice-free.
This distinction between the two processes is reflected in the temperature calculation, which uses a case-specific approach.

\subsubsection{Geometrical properties}

Basic geometric properties like height $ z $, volume $ V $ etc. are calculated as follows:
\begin{subequations}
\begin{align}
	 & V_\mathrm{hx,1} = \pi r_\mathrm{hx}^2 l_\mathrm{hx}, \label{eq:V_hx,1} \\
	 & V_\mathrm{hx,2} = \pi (r_\hx + \delta_\hx)^2 l_\mathrm{hx}, \label{eq:V_hx,2}   \\        
	 & m_\mathrm{hx} = \frac{\rho_{\hx} V_{\hx,1}}{n_\mathrm{hx} n_\w}, \label{eq:m_hx}                  \\
	 & V_\mathrm{is} = V_{\w} + V_{\hx,2}, \label{eq:V_is}                                  \\
	 & z_\w = \frac{V_\mathrm{is}}{\pi r_{\w}^2 n_{\w}}, \label{eq:h_is}                      \\
	 & m_{\w} = \frac{\rho_{\w} V_{\w}}{n_{\w}} \label{eq:m_w}, \\
	 & A_{\w} = \pi r_\w^2. \label{eq:A_w}
\end{align}
\end{subequations}
The volume of the medium inside the heat exchanger $ V_{\mathrm{hx},1} $ is calculated according to \eqref{eq:V_hx,1}, where $ r_\mathrm{hx} $ and $ l_\mathrm{hx} $ are the inner radius and total length of the heat exchanger pipes, respectively.
The total volume of the heat exchanger coils $ V_{\mathrm{hx},2} $ is defined by \eqref{eq:V_hx,2}, with $ \delta_\hx $ being the thickness of the heat exchanger pipes.
The mass of contained network medium per heat exchanger volume and water layer $ m_\hx $ is defined by \eqref{eq:m_hx}, with $ \rho_\hx $ being the density of the medium inside the heat exchanger.
The volume of the ice storage $ V_\mathrm{is} $ \eqref{eq:V_is} is equal to the sum of the water volume $ V_{\w} $ and heat exchanger volume $ V_{\mathrm{hx},2} $.
The height per water volume $ z_\w $ is then determined by rearranging the formula for the volume of a cylinder \eqref{eq:h_is}, where $ r_{\w} $ is the radius of a water layer.
The mass per layer $ m_\w $ is defined by \eqref{eq:m_w}, with the density of water $ \rho_\w $, and the cross-sectional area $ A_\w $ of the water layers given by \eqref{eq:A_w}. 

During the extraction phase, the water eventually starts to freeze and ice is forming around the extraction heat exchanger pipes.
The radius of the ice $ r_{\ice,i} $ around the heat exchanger pipes and the volume fraction of ice $ \phi_{\ice,i} $ are described by 
\begin{subequations}
	\begin{align}
		r_{\ice,i} = & \sqrt{\frac{V_\w \phi_{\ice,i}(t)}{n_\w \pi l_\hx} + (r_\hx+\delta_{\hx})^2}, \label{eq:r_ice} \\
		\phi_{\ice,i}(t) = & \mathrm{min}\biggl(\mathrm{max}\Bigl(0, \frac{T_{\w,i}(t)}{T_{\w,\mathrm{ice}}}\Bigr),1\biggr), \label{eq:phi_ice}
	\end{align}
\end{subequations}
for $ i = 1, \dots n_{\w} $, with $ T_{\w,\mathrm{ice}} =  \SI{-1}{\degreeCelsius} $. 

The volume $ V_{\co,i} $ and the mass $ m_{\co,i} $ of the concrete wall of the storage are calculated using
\begin{align}
	V_{\co,i} & =
	\begin{cases}
		\pi \bigl((r_{\w} + \delta_{\co})^2 - r_{\w}^2\bigr) z_{\w} + \pi r_{\w}^2 \delta_{\co}, & \text{if } i=1 \mathrm{\ or\ } i=n_{\w}, \\
		\pi \bigl((r_{\w} + \delta_{\co})^2 - r_{\w}^2\bigr) z_{\w} & \text{otherwise,}
	\end{cases} \\
	m_{\co,i} &= \rho_\co V_{\co,i},
\end{align}
for $ i = 1, ..., n_{\w} $, where $ \delta_{\co} $ is the thickness of the concrete walls and $ \rho_\co $ the density of concrete.
The top and bottom part of the concrete shell have additional parts for the base and lid of the storage.

The soil around the ice storage is horizontally and vertically discretized into $ n_{\s} $ layers of hollow cylinders around the storage.
The volume of the $ n_{\s} $ soil layers $ V_{\s,i,j} $ with $ i=1,\dots, n_{\w} $ and $ j=1,\dots,n_{\s} $ is given by
\begin{subequations}
\begin{align}
	\begin{split}
	V_{\s,i,j} = &\pi \Bigl( (r_{\w} + \delta_{\co} + j \delta_{\s})^2 - \bigl(r_{\w} + \delta_{\co} + (j - 1) \delta_{\s}\bigl)^2 \Bigr) z_{\w} \\
	&+ \pi r_{\w}^2 \delta_{\s}, \quad \text{if } i=1 \mathrm{\ or\ } i=n_{\w},
	\end{split} \\
	V_{\s,ij} = &\pi \Bigl( (r_{\w} + \delta_{\co} + j \delta_{\s})^2 - \bigl(r_{\w} + \delta_{\co} + (j - 1) \delta_{\s}\bigl)^2 \Bigr) z_{\w}, \\ &\textrm{otherwise}. \nonumber
\end{align}
\end{subequations}

\subsubsection{Heat exchanger}

The temperatures of the heat exchanger coils $ T_{\hx,i} $ are determined by
\begin{subequations}
\begin{align}
	\frac{\dn T_{{\hx},i}(t)}{\dt} = &\frac{\dot{m}_{\hx}(t) c_\f\Bigl(T_{{\hx},i-1}(t) - T_{{\hx},i}(t)\Bigr) - \dot{Q}_{{\w,\hx},i}(t)}{m_{\hx} c_{\f}}, \label{eq:Thx} \\
	\dot{Q}_{{\w,\hx},i}(t) = & U A \Bigl( T_{{\hx},i}(t) - T_{{\w},i}(t) \Bigr),  \label{eq:Qdot_hx}
\end{align}
\end{subequations}
for $ i=1, \dots, n_\w $.
For the first part of the heat exchanger coils, the temperature $ T_{\hx,0} $ refers to the inlet temperature $ T_{\n,\ret} $ (see Fig.~\ref{fig:is_bp}).
The heat flow $ \dot{Q}_{\w,\hx,i} $ between the heat exchanger and the respective water layer is described by \eqref{eq:Qdot_hx}, where $ U $ is the thermal transmittance.
Depending on the heat exchanger use, the term $ U A_i $ is described by
\begin{align}
	UA_i =
	\begin{cases}
		\frac{2 \pi l_\hx/n_\hx/n_\w}{\frac{1}{\alpha_{\f,\hx} r_\hx} + \frac{1}{\lambda_\tu}\ln \Bigl(\frac{r_{\hx} + \delta_{\hx}}{r_{\hx}} \Bigr) + \frac{1}{\lambda_\ice}\ln \Bigl(\frac{r_{\ice,i}}{r_{\hx} + \delta_{\hx}} \Bigr) + \frac{1}{\alpha_{\hx,\w} r_\hx}},                & \text{if \ case\ 1},         \\
		\frac{2 \pi l_\hx/n_\hx/n_\w}{\frac{1}{\alpha_{\f,\hx} r_\hx} + \frac{1}{\lambda_\tu}\ln \Bigl(\frac{r_{\hx} + \delta_{\hx}}{r_{\hx}} \Bigr) + \frac{1}{\alpha_{\hx,\w} r_\hx}}, & \text{otherwise,}
	\end{cases}
\end{align}
for $ i = 1,\dots, \nw $.
Case 1 indicates that the extraction heat exchanger is used, otherwise the regeneration heat exchanger is used.
At the inner wall of the pipes, the heat transfer coefficient is $ \alpha_{\f,\hx} $.
The heat conduction through the heat exchanger pipe and any ice layer, with radius $ r_{\ice,i} $, is influenced by the thermal conductivities $ \lambda_{\tu} $ and $ \lambda_\ice $, respectively.
The heat transfer coefficient on the outer wall of the heat exchanger or ice surface corresponds to $ \alpha_{\hx,\w} $.
It is assumed that the same temperatures apply to all $ n_{\hx} $ heat exchanger coils per water layer. 

\subsubsection{Water}
For the energy balance of the water layers, a heat exchange with the concrete walls of the ice storage and between the layers must also be considered.
The temperatures of the water volumes $ T_{\w,i} $ are determined by
\begin{subequations}
	\begin{align}
		\frac{\dn T_{\w,1}(t)}{\dt} = & \frac{n_{\hx} \dot{Q}_{{\w,\hx},1}(t) - \dot{Q}_{\w,\co,1}(t) + \dot{Q}_{\nc,1}}{m_{\w} c_{\w,1}(T_{\w,1})}, \label{eq:Q_nc1} \\
		\frac{\dn T_{\w,i}(t)}{\dt} = & \frac{n_{\hx} \dot{Q}_{{\w,\hx},i}(t) - \dot{Q}_{\w,\co,i}(t) - \dot{Q}_{\nc,i-1} + \dot{Q}_{\nc,i}}{m_{\w} c_{\w,i}(T_{\w,i})}, \label{eq:Q_nci} \\
		& i = 2, \dots, \nw-1, \nonumber \\
		\frac{\dn T_{\w,\nw}(t)}{\dt} = & \frac{n_{\hx} \dot{Q}_{{\w,\hx},\nw}(t) - \dot{Q}_{\w,\co,\nw}(t) - \dot{Q}_{\nc,\nw-1}}{m_{\w} c_{\w,\nw}(T_{\w,\nw})}.
	 	\label{eq:Q_ncnw}
	\end{align}
\end{subequations}
The heat flow between the water and the concrete wall, $ \dot{Q}_{\w,\co,i} $, as well as the natural convection within the layers, $ \dot{Q}_{\nc,i} $, influence the temperatures of the water volumes.
Note, that the specific heat capacity $ c_{\w,i}(T_{\w,i}) $ depends on the temperature $ T_{\w,i} $ and hence the state of aggregation of the water.
It can be determined by
\begin{align}
	c_{\w,i}(T_{\w,i}) =
	\begin{cases}
		c_{\w},                & \text{if } T_{\w,i}(t) > T_{\w,\f},         \\
		c_{\mathrm{ice}},          & \text{if } T_{\w,i}(t) < T_{\w,\mathrm{ice}}, \\
		\Delta h_{\fus}, & \text{otherwise,}
	\end{cases}
\end{align}
for $ i = 1, ... ,n_{\w} $.
It is assumed that a water layer starts to freeze at $ T_{\w,\f} = \SI{0}{\degreeCelsius} $ and is completely frozen at $ T_{\w,\mathrm{ice}} = \SI{-1}{\degreeCelsius} $.
This is based on the assumption that the crystallization process starts at the pipes of the heat exchanger and propagates towards the walls.
Therefore, the pipes may already be covered with a layer of ice and an increasing block of ice, while the outer part of the water layer is still liquid.
During the crystallization process, the specific heat capacity equals the normalized specific enthalpy of fusion, $ \Delta h_\mathrm{fus} = \SI{333.55}{\frac{\kilo\joule}{\kilogram\kelvin}} $, since the crystallization of water is assumed to occur between \SI{0}{\degreeCelsius} and \SI{-1}{\degreeCelsius}, this applies to a temperature range of \SI{1}{\kelvin}.
This is about 80 times the value of the heat capacity of water ($ c_{\w} = \SI{4.182}{\frac{\kilo\joule}{\kilogram\kelvin}}$) and 160 times the value of the heat capacity of ice ($ c_\mathrm{ice} = \SI{2.1}{\frac{\kilo\joule}{\kilogram\kelvin}}$) for one Kelvin \cite{Kuchling2022}.

The natural convection within the storage is given by
\begin{align}
	\dot{Q}_{\nc,i}(t)  = \lambda_{\w} \frac{A_{\w}}{z_\w} \Bigl(T_{\w,i+1}(t) - T_{\w,i}(t)\Bigr), \quad i=1,\dots,n_\w-1, \label{eq:Qdot_nc} 
\end{align}
where $ \lambda_{\w} $ is the thermal conductivity of water. 
The heat flow between the layers of water and concrete walls $ \dot{Q}_{\w,\co,i} $ are determined by
\begin{subequations}
	\begin{align}
		\begin{split}
			\dot{Q}_{\w,\co,1}(t) = & \frac{\pi}{\frac{1}{(2 r_\w z_\w + r_w^2)\alpha_{\w,\co}} + \frac{1}{2 \lambda_{\co} z_\w} \ln\Bigl(\frac{r_\w+\delta_{\co}/2}{r_\w}\Bigr) + \frac{\delta_{\co}/2}{\lambda_{\co} r_w^2}} \\ 
			& \cdot \Bigl(T_{\w,1}(t) - T_{\co,1}(t)\Bigr), \label{eq:Qdot_wc1} \\
		\end{split}\\
		\begin{split}
			\dot{Q}_{\w,\co,i}(t) = & \frac{2 \pi z_\w}{\frac{1}{\alpha_{\w,\co} r_\w} + \frac{1}{\lambda_{\co}} \ln\Bigl(\frac{r_\w+\delta_{\co}/2}{r_\w}\Bigr)} \Bigl(T_{\w,i}(t) - T_{\co,i}(t)\Bigr), \\
			& i=2,\dots,\nw-1, \label{eq:Qdot_wci}  \\
		\end{split} \\
		\begin{split}
			\dot{Q}_{\w,\co,\nw}(t) = & \frac{\pi}{\frac{1}{(2 r_\w z_\w + r_w^2)\alpha_{\w,\co}} + \frac{1}{2 \lambda_{\co} z_\w} \ln\Bigl(\frac{r_\w+\delta_{\co}/2}{r_\w}\Bigr) + \frac{\delta_{\co}/2}{\lambda_{\co} r_w^2}} \\ 
			& \cdot \Bigl(T_{\w,\nw}(t) - T_{\co,\nw}(t)\Bigr), 
			\label{eq:Qdot_wcnw} \\
		\end{split}
	\end{align}
\end{subequations}
for $ i = 1, \dots n_{\w} $, with $T_{\co} $, $ \alpha_{\w,\co} $ and $ \lambda_{\co} $ being the temperature of the concrete walls, the heat transfer coefficient and heat conductivity of concrete and $ \delta_{\co} $ the thickness of the storage shell.
The heat flow equations contain a heat convection component and a heat conduction term from the inner wall of the concrete shell to the center of the shell, as the temperatures are calculated in the center of the finite volumes.

\subsubsection{Concrete walls}
The temperatures $ T_{\co,i} $ of the concrete walls are described by
\begin{align}
	\frac{\dn T_{\co,i}(t)}{\dt} & = \frac{\dot{Q}_{\w,\co,i}(t) - \dot{Q}_{\co,\s,i}(t)}{m_{\co,i} c_{\co}} , \label{eq:T_c}
\end{align}
for $ i,\dots,n_\w $, where $ c_\co $ is the specific heat capacity of concrete.
The temperature of the concrete shell is directly influenced by the adjacent water layers via $ \dot{Q}_{\w,\co,i} $ and soil layer via $ \dot{Q}_{\co,\s,i} $.

The heat transfer between the concrete hull and surrounding soil is determined by
\begin{subequations}
	\begin{align}
		\dot{Q}_{\co,\s,1}(t) = & \frac{\pi}{\frac{1}{2 \lambda_{\co} z_\w} \ln\Bigl(\frac{r_\w+\delta_{\co}}{r_\w+\delta_{\co}/2}\Bigr) + \frac{1}{2 \lambda_{\s} z_\w} \ln\Bigl(\frac{r_\w+\delta_{\co}+\delta_{\s}/2}{r_\w+\delta_{\co}}\Bigr) +  \frac{\delta_{\co}/2}{\lambda_{\co} r_w^2} +  \frac{\delta_{\s}/2}{\lambda_{\s} r_w^2}} \nonumber \\ 
		& \cdot \Bigl(T_{\co,1}(t) - T_{\s,1,1}(t)\Bigr), \label{eq:Qdot_cs1}\\
		\dot{Q}_{\co,\s,i}(t)  = & \Bigl(\frac{2 \pi z_{\w}}{\frac{1}{\lambda_{\co} }\ln\Bigl(\frac{r_{\w} + \delta_{\co}}{r_{\w} + \delta_{\co}/2} \Bigr) + \frac{1}{\lambda_{\s} }\ln\Bigl(\frac{r_{\w} + \delta_{\co} + \delta_{\s}/2}{r_{\w} + \delta_{\co}} \Bigr)} \Bigr) \Bigl(T_{\co,i}(t) - T_{\s,i,1}(t)\Bigr) \label{eq:Qdot_csi},  \nonumber \\
		& i=2,\dots, \nw-1, \\
		\dot{Q}_{\co,\s,\nw}(t) = & \frac{\pi}{\frac{1}{2 \lambda_{\co} z_\w} \ln\Bigl(\frac{r_\w+\delta_{\co}}{r_\w+\delta_{\co}/2}\Bigr) + \frac{1}{2 \lambda_{\s} z_\w} \ln\Bigl(\frac{r_\w+\delta_{\co}+\delta_{\s}/2}{r_\w+\delta_{\co}}\Bigr) +  \frac{\delta_{\co}/2}{\lambda_{\co} r_w^2} +  \frac{\delta_{\s}/2}{\lambda_{\s} r_w^2}} \nonumber \\ 
		& \cdot \Bigl(T_{\co,\nw}(t) - T_{\s,\nw,1}(t)\Bigr). \label{eq:Qdot_csnw}
	\end{align}
\end{subequations}

It is assumed, that the concrete wall interacts with the first soil layer whose temperature is given by $ T_{\s,i,1} $. The first index $ i $ indicates the water layers, while the second index $ j $ reflects the soil layers.
In addition to the heat transfer for hollow cylinders through the sides of the storage \eqref{eq:Qdot_csi}, the heat transfer through the base and lid of the storage has additional terms for the heat transfer through the circular surfaces of the base and the lid \eqref{eq:Qdot_cs1}, \eqref{eq:Qdot_csnw}.

\subsubsection{Soil}

For modeling the soil layers surrounding the ice storage, $ n_{\s} $ temperatures for every water layer $ n_{\w} $ are considered, i.e. the ground is discretized along the horizontal and vertical axis of the ice storage, see Fig.~\ref{fig:is_dic}.
Hence, $ n_{\w} n_{\s} $ soil temperatures are calculated.
The temperatures $ T_{\s,i,j} $ are given by
\begin{subequations}
\begin{align}
	\frac{\dn T_{\s,i,1}(t)}{\dt}       = &\frac{\dot{Q}_{\co,\s,i}(t) - \dot{Q}_{\s,i,1}(t)}{V_{\s,i,1}C_{\s,i,1}} , \label{eq:Ts1} \\
	\frac{\dn T_{\s,i,j}(t)}{\dt}        = &\frac{\dot{Q}_{\s,i,j-1}(t) - \dot{Q}_{\s,i,j}(t)}{V_{\s,i,j}C_{\s,i,j}} ,  \quad j=2,\dots,n_{\s}, \label{eq:Ts}
\end{align}
\end{subequations}
for $ i=1,\dots,\nw$.
The temperature of the very first soil layer is determined by \eqref{eq:Ts1}, influenced by the heat flow from the concrete wall $ \dot{Q}_{\co,\s,i} $ and the heat flow between the neighboring soil layer $ \dot{Q}_{\s,i,1} $.
For the second up to the last layer, the soil temperature is influenced by the heat flow between the respective soil layer and its previous and subsequent layer \eqref{eq:Ts}.
It is assumed that there is a universal temperature $ T_{\s,n_\s+1} $ at the boundaries of the considered soil, which corresponds to the uninfluenced soil temperature (cf. \eqref{eq:T_s_bd}) at the installation depth.
The volumetric heat capacity of the soil, $ C_{\s,i,j} $ is calculated according to \eqref{eq:C_s}.

The propagating heat flow within the soil layers and the boundary is determined by
\begin{subequations}
	\begin{align}
	\begin{split}
		\dot{Q}_{\s,1,j} = &\Bigl(\frac{2 \lambda_{\s} \pi z_{\w}}{\ln\Bigl(\frac{r_{\w} + \delta_{\co} + (j+1/2)\delta_{\s}}{r_{\w} + \delta_{\co} + (j - 1/2) \delta_{\s}}\Bigr)} + \frac{\lambda_{\s}}{\delta_{\s}} \pi r_{\w}^2 \Bigr)  \\
		&\cdot \Bigl(T_{\s,1,j}(t) - T_{\s,1,j+1}(t)\Bigr), \quad j=1,\dots,n_{\s}, \label{eq:Qdot_s1}
	\end{split} \\
	\dot{Q}_{\s,i,j}          = &\frac{2 \lambda_{\s} \pi z_{\w}}{\ln\Bigl(\frac{r_{\w} + \delta_{\co} + (j+1/2)\delta_{\s}}{r_{\w} + \delta_{\co} + (j - 1/2) \delta_{\s}}\Bigr)} \label{eq:Qdot_si} \numberthis \\&\cdot \Bigl(T_{\s,i,j}(t) - T_{\s,i,j+1}(t) \Bigr), \quad i=2,\dots,n_{\w}-1, j=1,\dots,n_{\s},  \nonumber   \\
	\begin{split}
	\dot{Q}_{\s,\nw,j}           = &\Bigl(\frac{2 \lambda_{\s} \pi z_{\w}}{\ln\Bigl(\frac{r_{\w} + \delta_{\co} + (j+1/2)\delta_{\s}}{r_{\w} + \delta_{\co} + (j - 1/2) \delta_{\s}}\Bigr)} + \frac{\lambda_{\s}}{\delta_{\s}} \pi r_{\w}^2 \Bigr) \\
	&\cdot \Bigl(T_{\s,n_\w,j}(t) - T_{\s,n_\w,j+1}(t) \Bigr), \quad j=1,\dots,n_{\s}. \label{eq:Qdot_snw}
	\end{split}                                                                        
	\end{align}
\end{subequations}

The heat flow between neighboring layers is described by the heat flow of cylindrical shapes \eqref{eq:Qdot_si}.
The soil layers underneath and above the ice storage  have additional terms for the circular parts \eqref{eq:Qdot_s1}, \eqref{eq:Qdot_snw}.

\subsection{Consumer heat exchanger (at heat transfer stations)}
\begin{figure}[t!]
	\centering
	\includegraphics[width=0.5\textwidth]{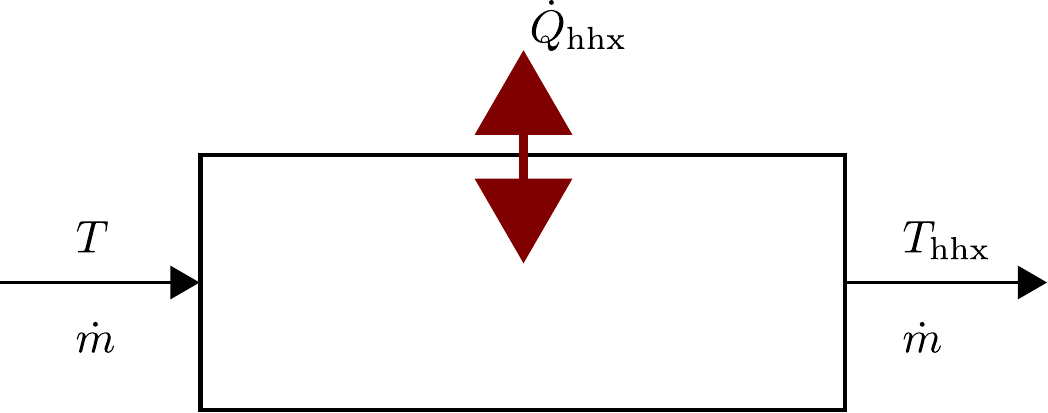}
	\caption{Schematic depiction of the heat exchanger at the heat transfer station.}
	\label{fig:hx}
\end{figure}

The connected buildings within \acrshort{5GDHC} networks are often supplied with heat via individual heat transfer stations that function as a heat exchanger between the network side and the consumer side (see Fig.~\ref{fig:hx}).
This separates the network thermohydraulically from the individual buildings.
The heat exchangers in the transfer stations generally operate in counterflow. 
For the network operator, the heat exchanger may serve as a billing point for heat consumption. 
The heat exchangers for the individual consumers are therefore modeled to cover the network side and only consider a heat demand curve on the building site.
Hence, the relevant part of the heat transfer station model is the primary side of the heat exchanger and its temperature development
\begin{align}
	\frac{\dn T_\hhx(t)}{\dt} &= \frac{\dot{m}_\hhx(t) c_{\f} \Bigr(T(t) - T_\hhx(t)\Bigl) + \dot{Q}_\hhx(t)}{m_\hhx c_{\f}} , \label{Thhx}
\end{align}
where $ m_\hhx $ and $ c_{\f} $ are the mass of the fluid inside the primary side of the heat exchanger and the specific heat capacity of the fluid, respectively, $ T $ the temperature of the medium at the inlet and $ \dot{Q}_\hhx $ the heat flow requested from the buildings.

\subsection{Circulation pump}

A central pump is used to regulate the mass flow in order to maintain the pressure in the network. 
The power consumption of the pump is described by the following relationship:
\begin{align}
	P_\mathrm{el}(t) = \frac{\Delta p \dot{m}_\n}{\rho_\f \eta}, \label{eq:Pel}
\end{align}
where $ \Delta p $ is the total pressure loss of the medium running through the network pipes, $ \rho_\f $ the density of the medium, and $ \eta $ the efficiency of the pump.

\section{Comparison of simulation results and measurements}\label{secResults}

In this section, the setup for the simulation study of the heating network in Gutach-Bleibach is introduced followed by the results of the simulation study. 
The results are then evaluated using \acrfull{ASHRAE} Guidelines \cite{ASHRAE2014} to determine the model validation using the \acrfull{cvrmse} and \acrfull{nmbe}.

\subsection{Simulation study}

\begin{figure}[t]
	\centering
	\includegraphics[width=0.8\textwidth]{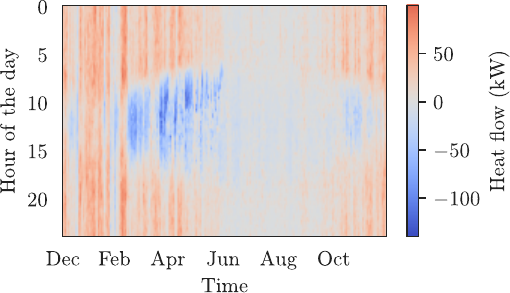}
	\caption{Total heat flow at the transfer stations of the consumers for every day of the year and every hour of the day. Positive values refer to the network supplying heat to the buildings.}
	\label{fig:Qdot_h}
\end{figure}

\begin{figure*}[t]
	\centering
	\includegraphics[width=\textwidth]{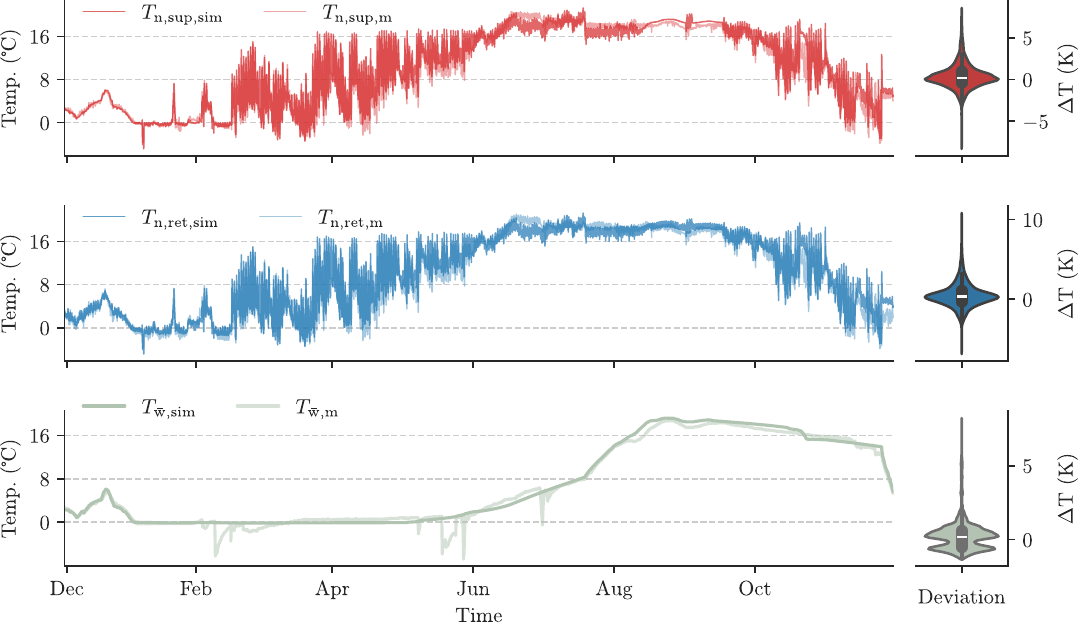}
	\caption{Progression of the supply $ T_\mathrm{n,sup} $ and return temperature $ T_\mathrm{n,ret} $ of the heating network, as well as the average water temperature in the ice storage $ T_\mathrm{\bar{w}} $.  In the right hand side, violin plots are used to show the deviations from measurements (m) and values of the simulation (sim).}
	\label{fig:T_n}
\end{figure*}

The models introduced are used for a simulation study on the \acrshort{5GDHC} network in Gutach-Bleibach, Germany, as described in Section~\ref{secSystem}.
For this task, the models are implemented in the Modelica Language \citep{Fritzson1998}.
An open-source implementation of the model is provided at \href{https://github.com/hka-esa/n5GDHC_sim}{n5GDHC\_sim} \cite{Git5GDHC}.
Modelica is an object-oriented modeling language for modeling cyber-physical systems with differential, algebraic and discrete equations.
The district network model consists of several individual sub-models that are combined into larger models. 
For example, the pipe model is used for the supply and return sections of a network part and supplemented with the model of the surrounding soil.
The basic dynamics of the pipe model are also used for the pipes running through the ice storage. 

OpenModelica was used as the simulation environment for the dynamic simulations \citep{Fritzson2020}.
A temporal discretization is carried out in order to simulate the temporal development of the variables.
This temporal discretization is done with the help of integrators that are available in OpenModelica.

The network is simulated with 30 buildings connected to the ice storage via house transfer stations and pipe connections.
The simulation is carried out for about one year of operation and is later compared to available measurements of the network.

The control strategy of the ice storage as described in Section~\ref{subsec:CS} is carried out using a PI controller to control the (partial) activation of the ice storage depending on the seasonal reference temperature for the supply temperature of the network.

The inlet and outlet temperatures of the medium at the house transfer stations, including the mass flow, are available as measured data for the network. 
The heat outputs requested by the consumers were calculated from this data. 
These are used as input data $ \dot{Q}_\hhx $ in \eqref{Thhx} for the simulation. 
The cumulative heat energy transferred at the transfer stations is shown in Fig.~\ref{fig:Qdot_h}.
The heat exchangers are operated with a specified temperature spread $ \Delta T_\hhx $ between the flow and return so that the mass flow through the heat exchangers in \eqref{Thhx} is calculated by:
\begin{align}
	\dot{m}_\hhx(t) = \frac{| \dot{Q}_\hhx (t) |}{c_\f \Delta T_\hhx}.
\end{align}
Material properties of the installed components on site and thermophysical parameters are taken from data sheets or literature like \acrfull{DIN} standards and \cite{Kuchling2022}.

\subsection{One year simulation study}

The simulation model presented will be tested over a period of around one year and compared with corresponding measurement data.
The available measurement data are temperatures of the fluid within the network station near the ice storage and the water temperatures inside the ice storage. 
At the time of the measurements, 30 buildings were connected to the network.
The simulation starts during the heating season in December and runs until the end of November the following year.

The connected buildings act as both heat consumers and heat producers in the network.
Based on the cumulative heat flow, it can be seen that there is significantly more activity and exchange with the grid during the cold season and transition periods (see Fig.~\ref{fig:Qdot_h}).
This is presumably also due to the sometimes greater temperature differences between the building and grid side.
Positive values indicate that heat is being extracted, negative values indicate that heat is being fed into the grid.

The simulation results and their comparison with the measured values for selected temperatures are shown in Fig.~\ref{fig:T_n}.
In the upper plot, the supply temperature of the network calculated from the simulation (sim) is compared with the corresponding measured values (m).
The violin plot next to it shows the deviations between both values.
The middle plot shows the return temperatures of the network, and the lower plot shows the average water temperature in the ice storage.

\begin{figure*}[t!]
	\centering
	\includegraphics[width=0.95\textwidth]{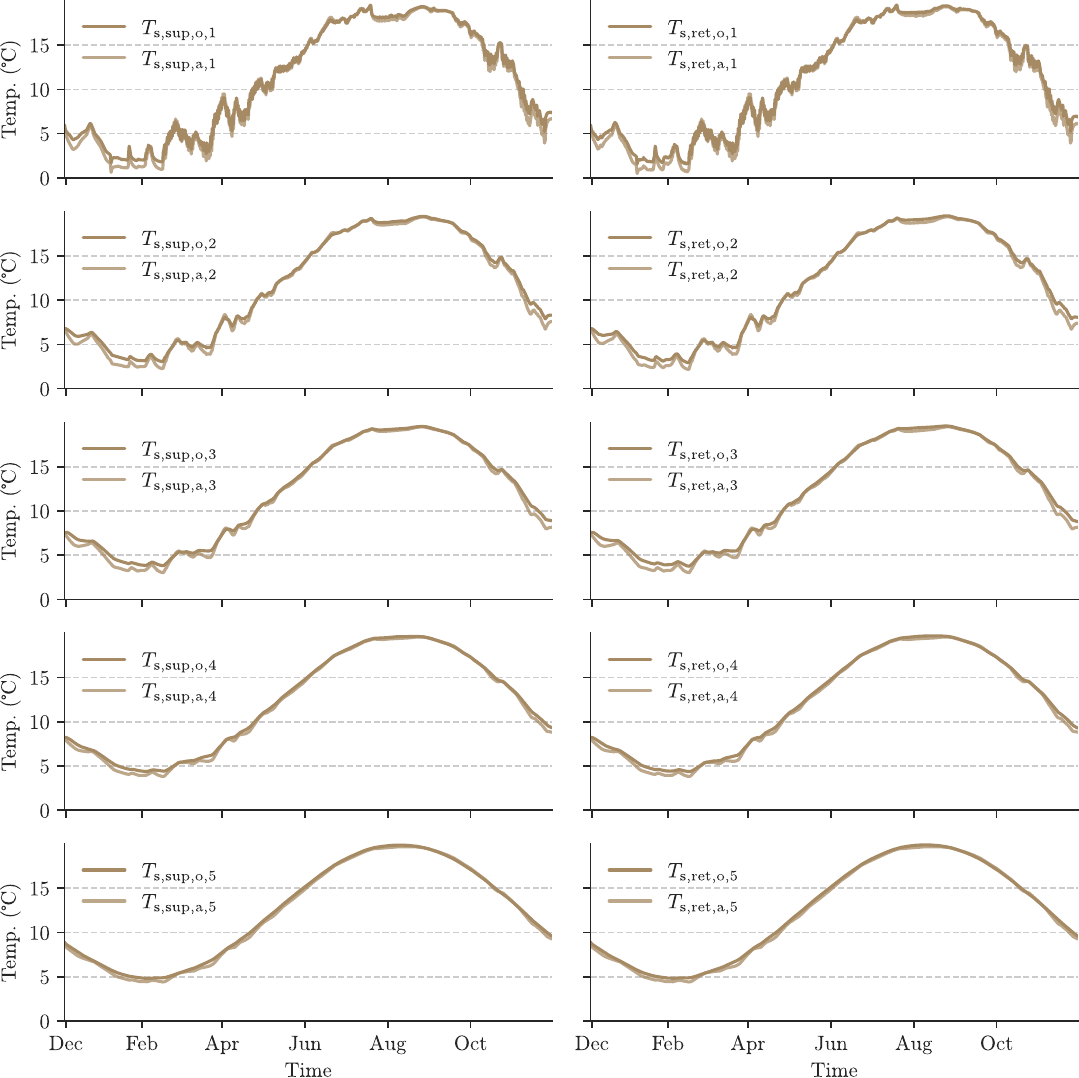}
	\caption{Soil temperatures of an exemplary pipe section of the supply side throughout the simulation for $ n_\s = 5 $. The plots show the temperature development for both, the outer and adjacent, parts of the supply and return side and for every soil layer $ i=1,\dots, n_\mathrm{s} $.}
	\label{fig:T_s}
\end{figure*}

At the start of the simulation, the ice storage is in full operation and is used to supply heat to the network.
It can be seen that the storage eventually begins to freeze at the beginning of January, so that a relatively constant supply temperature of $ \SI{0}{\degreeCelsius} $ is reached in the network.
The ice storage is then used until around the end of February, when it is fully switched to bypass and no longer interacts with the network.
Up to this point, the values show a good match between simulation and measurement.
From the beginning of June, the grid temperatures slowly reach their maximum values around $ \SI{20}{\degreeCelsius} $ and the ice storage is used to cool the buildings.
Regeneration of the ice storage also causes the water temperature in the storage to rise.
The measured water temperatures show several downward peaks between February and July, which can be attributed to measurement errors and faulty instrumentation.
This results in a strong stretching of the violin plot.
Towards the end of the simulation period and with the start of the heating period, the network temperatures fall, and the ice storage is used again for heating purposes.
The seasonal variation of temperatures within the network can also be seen in Fig.~\ref{fig:T_s}, which shows the soil temperatures around a sample section of pipe.
While the temperatures in all surrounding soil layers show a similar pattern, a higher volatility of the temperatures can be observed in the layers close to the pipes due to the stronger influence of the medium temperature in the pipes.

\subsubsection{Model validation}
Model validation is determined using \acrshort{ASHRAE} Guideline 14, which states that a model is validated if it is within \SI{\pm10}{\percent} and \SI{30}{\percent} for the \acrshort{nmbe} and the \acrshort{cvrmse}, respectively.
The model is therefore considered to be validated, as shown by the values for the available measurements in Table \ref{table_example}.
For the \acrshort{nmbe}, the range is between \SI{3.93}{}-\SI{5.00}{\percent} and for the \acrshort{cvrmse} between \SI{14.57}{}-\SI{16.78}{\percent} for all measured values, which are within the required limits.

\begin{table}[h]
	\caption{Statistical evaluation of the simulation results and comparison with measured values according to \acrshort{ASHRAE} Guidelines \cite{ASHRAE2014} and the key figures \acrshort{cvrmse} and \acrshort{nmbe}.}
	\label{table_example}
	\begin{center}
		\begin{tabular}{lrrrr}
			\toprule
			 & $ T_{\n,\sup} $ & $ T_{\n,\ret} $ & $ T_{\bar{\w}} $ & Limit \\
			\cmidrule(lr){1-5}
			NMBE (\%):   & \SI{3.93}{} & \SI{5.00}{} & \SI{4.54}{} & $ \pm \SI{10}{} $  \\
			CVRMSE (\%): & \SI{16.33}{} & \SI{16.78}{} & \SI{14.57}{} & \SI{30}{} \\
			\bottomrule
		\end{tabular}
	\end{center}
\end{table}

\section{Conclusion and future work}\label{secDiscussion}
In this work, the modeling of uninsulated pipe and its surrounding soil, an ice storage and heat transfer stations as used in \acrshort{5GDHC} networks were presented.
The models are formulated in an software-independent way and were implemented in the Modelica language to be verified and validated in a simulation study.
The individual models were combined to form a model for a \acrshort{5GDHC} network in southern Germany in order to carry out a validation using available measured data.
The network utilizes the ice storage as a \acrshort{ltes} to supply heat and cold throughout the year to 30 buildings.
The model presented emphasizes the spatially and temporally resolved thermal relationships of a low-temperature network. 
The network model was simulated in an experimental study over a period of one year and compared with available temperatures at measurement points in the network.

The study shows that the model provides a robust representation of network dynamics throughout the year, achieving good agreement with measured values.
Key temperature values such as supply and return temperature of the network and water temperature inside the storage have a maximum \acrshort{nmbe} of \SI{5.00}{\percent} and \acrshort{cvrmse} of \SI{16.78}{\percent}, which corresponds to a calibrated model in accordance with \acrshort{ASHRAE} Guidelines \cite{ASHRAE2014}. 

Influencing factors that have not yet been taken into account, such as different soil conditions at the boundary of the terrain under consideration, and the effects of energy input from precipitation, could be the subject of further work. 
However, the latter appears to be negligible, at least in terms of energy quantity.
The software-independent modeling makes the model widely applicable and potentially suitable for use in the field of optimization. 
Given other requirements for the models in the context of optimization, the extent to which a reduction in complexity and model size is possible may need to be investigated.
In particular, future work could explore the impact of simplifying the model architecture on the accuracy and efficiency of an optimization approach.
This is particularly relevant in the context of the case distinction in the equations introduced in this work, which may require advanced optimization methods.

\section*{CRediT authorship contribution statement}
\textbf{Manuel Kollmar:} Writing – original draft, Writing – review and editing, Visualization, Validation, Methodology, Investigation, Data curation, Software, Conceptualization. \textbf{Adrian B{\"u}rger:} Writing – review and editing, Conceptualization, Methodology, Validation, Software. \textbf{Markus Bohlayer:} Writing – review and editing, Methodology. \textbf{Angelika Altmann-Dieses:} Writing – review and editing, Supervision, Funding acquisition. \textbf{Marco Braun:} Supervision, Project administration, Funding acquisition. \textbf{Moritz Diehl:} Writing – review and editing, Supervision, Conceptualization.

\section*{Declaration of competing interest}
The authors declare that they have no known competing financial interests or personal relationships that could have appeared to influence the work reported in this paper.

\section*{Declaration of generative AI and AI-assisted technologies in the writing process}
During the preparation of this work the authors used ChatGPT in order to enhance language and readability. After using this tool/service, the authors reviewed and edited the content as needed and take full responsibility for the content of the published article.

\section*{Acknowledgement}
The authors kindly thank the Stadtwerke Bühl GmbH and the Bühler Bürger-Energiegenossenschaft for their support and cooperation in this work and Arnold Schmid from Innovativ SCHMID for providing the image of the inside of the ice storage.

This research was supported by the Carl-Zeiss-Stiftung via the project P2021-08-008 and by the Federal Ministry for Economic Affairs and Climate Protection via project 03EN3054B.

\section*{Data availability}
The model implemented in Modelica is openly available at: \href{https://github.com/hka-esa/n5GDHC_sim}{https://git\-hub.com/hka-esa/n5GDHC\_sim}.
The original data is confidential and therefore the authors do not have permission to share this data.

\label{}

\bibliographystyle{elsarticle-num-names}
\bibliography{5GDHCmodel}

\end{document}